\documentclass{article}

\usepackage{arxiv}

\usepackage[utf8]{inputenc} 
\usepackage[T1]{fontenc}    
\usepackage{hyperref}       
\usepackage{url}            
\usepackage{booktabs}       
\usepackage{amsfonts}       
\usepackage{nicefrac}       
\usepackage{microtype}      
\usepackage{lipsum}		
\usepackage{graphicx}
\usepackage{natbib}
\usepackage{doi}
\usepackage{fancyhdr}
\usepackage{times}
\usepackage{bm}
\usepackage{bbm}
\usepackage{amsmath, amssymb, enumitem}
\usepackage{tikz}
\usetikzlibrary{automata, positioning, arrows, shapes.geometric}
\usepackage{caption}
\usepackage{subcaption}
\usepackage{multirow}
\usepackage{adjustbox}
\usepackage{lscape} 
 \usepackage{rotating}
 \usepackage{tikz-cd}
 \numberwithin{equation}{section}
 \newcommand{\indep}{\perp \!\!\! \perp}
\usetikzlibrary{graphs,decorations.pathmorphing,decorations.markings} 
\usetikzlibrary{arrows}
\usepackage{lscape} 
\usetikzlibrary{decorations.markings}
\tikzset{negated/.style={
        decoration={markings,
            mark= at position 0.5 with {
                \node[transform shape] (tempnode) {$\backslash$};
            }
        },
        postaction={decorate}
    }
}
\newtheorem{theorem}{Theorem}

\title{Estimands and cumulative incidence function regression in clinical trials: some new results on interpretability and robustness}

\author{ Alexandra Bühler \\
    Statistics and Actuarial Science \\
	University of Waterloo \\
	Waterloo, Canada \\
	\texttt{abuhler@uwaterloo.ca} \\
	\And
	Richard J. Cook \\
	Statistics and Actuarial Science \\
	University of Waterloo \\
	Waterloo, Canada \\
	\texttt{rjcook@uwaterloo.ca} \\
	\And 
	Jerald F. Lawless \\
	Statistics and Actuarial Science \\
	University of Waterloo \\
	Waterloo, Canada \\
	\texttt{jlawless@uwaterloo.ca} \\
}

\date{}
\begin{document}
\maketitle

\begin{abstract}
Regression analyses based on transformations of cumulative incidence functions are often adopted when modeling and testing for treatment effects in clinical trial settings involving competing and semi-competing risks. Common frameworks include the Fine-Gray model and models based on direct binomial regression. Using large sample theory we derive the limiting values of treatment effect estimators based on such models when the data are generated according to multiplicative intensity-based models, and show that the estimand is sensitive to several process features. The rejection rates of hypothesis tests based on cumulative incidence function regression models are also examined for null hypotheses of different types, based on which a robustness property is established. In such settings supportive secondary analyses of treatment effects are essential to ensure a full understanding of the nature of treatment effects. An application to a palliative study of individuals with breast cancer metastatic to bone is provided for illustration.
\end{abstract}


\section{Introduction}
\begin{figure}[ht]
\begin{center}
\noindent
\begin{minipage}[t]{0.49\textwidth}
\centering
\includegraphics[scale=0.48]{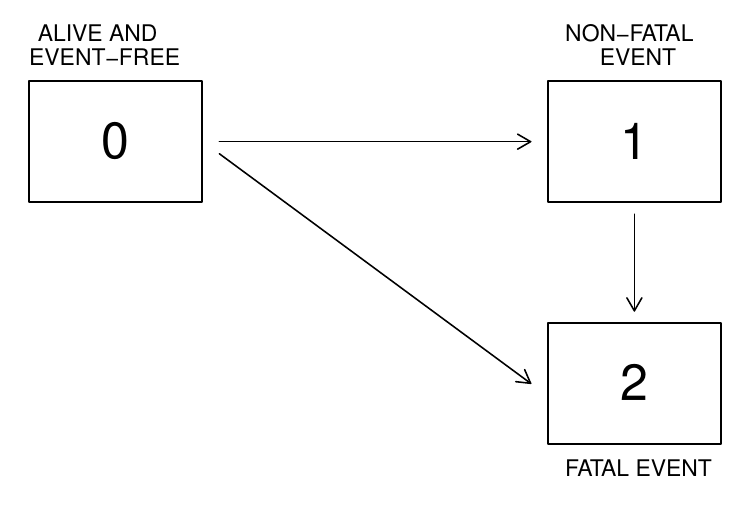} \\
{(a) A three-state illness-death process \newline}
\end{minipage}
\noindent
\begin{minipage}[t]{0.49\textwidth}
\begin{center}
\includegraphics[scale=0.48]{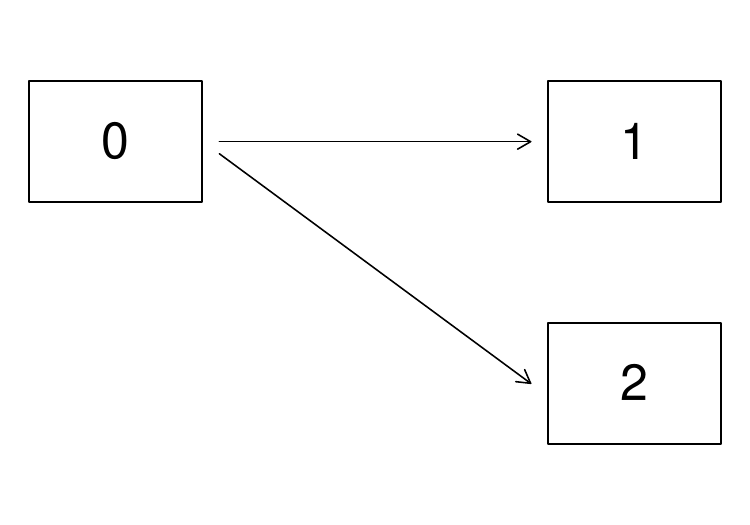} \\
 {(b) A three-state competing risks process \newline}
\end{center}
\end{minipage}
  \caption{Multistate diagrams for illness-death and competing risks processes.}
  \label{multistate-diagram-1}
\end{center}
\end{figure}

In clinical trials involving complex disease processes individuals are typically at risk of several types of events. In palliative trials of patients with metastatic cancer, individuals are at risk of cancer progression, skeletal metastases, related complications, and death \citep{Hortobagyi1996, Theriault1999}. In cardiovascular (CV) trials individuals may experience non-fatal myocardial infarctions, non-fatal strokes, CV-related death, and death from other causes \citep{Marso2016}. Multistate models offer a powerful framework for characterizing such processes \citep{Andersen1993, Cook2018, Buehler2022, Andersen2023}. \newline 
Intensity-based models for multistate processes are most closely aligned with how events unfold over time. When analyzing treatment effects in randomized trials however, intensity-based models induce time-dependent confounding by conditioning on the complete process history at a given time. They therefore do not support conventional causal inferences regarding the effect of randomized treatments \citep{Hernan2010, Aalen2015}, despite playing a crucial role in gaining deeper insights into disease processes. Because marginal models do not condition on post-randomization events they are more suitable in this context and have thus become the standard approach for evaluating the effects of a randomized treatment in clinical trials \citep{Buehler2022}. 
\newline 
Various types of marginal models have been discussed in the literature for different disease process settings \citep{Fine1999, Scheike2007, Scheike2008, Eriksson2015, Ghosh2002, Mao2016}, but a number of issues concerning the interpretability of treatment effects under model misspecification and robustness of tests have not been adequately addressed. We study the impact of particular kinds of model misspecification and focus on generalized linear models (GLMs) based on cumulative incidence functions (CIFs) in semi-competing risks settings (see Figure \ref{multistate-diagram-1}, \citealp{Gerds2012}).
We consider the relationship between the CIF for the non-fatal event and the process intensity functions in the spirit of \citet{Putter2020}, but we do so using large sample theory; see also \citet{Latouche2007} and \citet{Grambauer2010}. When the true data-generating mechanism is an illness-death process with proportional intensities, we derive the limiting values of two common estimators of treatment effect in the complementary log-log transformed CIF model considered by \citet{Fine1999}: one based on Fine and Gray's (FG) proposed estimation procedure \citep{Fine1999, He2016}, and the other based on so-called direct binomial (DB) regression \citep{Scheike2008}. The former procedure models the associated subdistribution hazard function through a weighted Cox-type approach, whereas the latter uses a weighted composite likelihood based on binomial estimating equations for estimation.
We investigate the dependence of limiting values of estimators on features of the underlying illness-death process, and find that the estimand arising from DB regression can be less sensitive than the estimand from the FG approach to variation in the true process. Such investigations give insights into the determinants of estimands which, along with supportive secondary analyses, can help in the interpretation of study results.
 \newline  
Tests for treatment effects are an essential component of randomized clinical trials. A standard approach to testing of no covariate effects on CIFs is the nonparametric log-rank-type test by \citet{Gray1988}; see also \citet{Poythress2020} for a recent discussion in the context of randomized trials. We argue however that tests in a trial's primary analysis should be related to corresponding estimands; here under the true intensity-based data generation process the implicit estimand is not $\beta$ but the limiting value $\beta^{\star}$ of the FG and DB treatment effect estimator as the sample size becomes arbitrarily large. Using asymptotic theory of misspecified models \citep{White1982, Struthers1986}, we examine whether FG- and DB-based Wald tests of no treatment effect are asymptotically valid under different hypotheses concerning the true illness-death process. We show that the tests control the type I error under two key hypotheses, provided they employ robust variance estimates.
\newline 
The remainder of this article is organized as follows. In Section \ref{sec2.1} we introduce notation for the illness-death process and define intensity functions. In Section \ref{sec-2.2} generalized linear models are defined for CIFs, following which we give estimating functions for the effect of a binary covariate for the FG approach and DB regression in Section \ref{sec3.1}. In Section \ref{sec-numericalstudies-CR} we study the limiting values of the estimators arising from FG and DB estimation as a function of the parameters of an intensity-based data-generating process. Section \ref{sec3.3} is concerned with model checking and goodness-of-fit tests. In Section \ref{sec-4.1} we consider tests for treatment effects and carefully outline several explicit hypotheses that may be considered in regulatory settings. Theoretical and numerical results concerning the effects of model misspecification on properties of FG- and DB-based Wald tests are provided in Section \ref{simulations-testing}. In Section \ref{sec5} we present an illustrative analysis of a trial of the effect of a bisphosphonate on fracture risk in breast cancer patients with skeletal metastases. Concluding remarks are given in Section \ref{sec6}. Some derivations and additional numerical results are provided in the Appendix and Online Supplementary Material. 

\section{Generalized linear models for marginal process features}
\label{sec2}
\subsection{Notation and illness-death processes}
\label{sec2.1}
We consider the setting of a two-arm phase III randomized clinical trial, and let $X$ be a binary covariate equal to $1$ for individuals in the experimental treatment group and $0$ otherwise. We assume the disease process can be represented by an illness-death model with state space $\mathcal{S}=\{0,1,2\}$; see Figure \ref{multistate-diagram-1}(a). In oncology trials, for example, state $0$ may represent the state occupied at the time of treatment assignment, with state $1$ entered upon occurrence of a non-fatal event (e.g. progression, relapse), and state $2$ entered upon death. We let $Z(t)$ denote the state occupied by an individual at time $t\geq0$, and assume
that the process $\{Z(t),t\geq 0\}$ starts in state $0$ at $t=0$. The intensity function for a $k - l$ transition at time $t$ is then defined as 
\begin{align}
\label{formula-intensity}
\lim_{\Delta t \downarrow 0} \dfrac{P(Z(t + \Delta t^{-}) = l \ | \ Z(t^{-})=k,  \mathcal{H}(t))}{\Delta t} = \lambda_{kl}(t | \mathcal{H}(t))\; , \ \ \ \ \ \ \ \ \ \ \ \ \ k,l \in \mathcal{S},  k < l
\end{align}
where the process history $\mathcal{H}(t)=\{Z(u), 0 \leq u < t, X \}$ contains information on the occurrence and timing of transitions over $[0,t)$ and the treatment indicator $X$. We consider Markov processes in which the transition intensities only depend on time $t$ and the state occupied at $t^-$, so $\lambda_{kl}(t | \mathcal{H}(t))=Y_k(t) \lambda_{kl}(t | X)$ where $Y_{k}(t)=\mathbbm{1}(Z(t^{-})=k)$ indicates that state $k$ is occupied at $t^{-}$. 
The set of all intensities (\ref{formula-intensity}) fully specifies the process, and any marginal features are functionals of the intensities \citep{Andersen2012.1}. We next formulate generalized linear models (GLMs) for some particular marginal features of an illness-death process. 

\subsection{Marginal process features and model formulation}
\label{sec-2.2}
With illness-death processes, it is common to conduct analyses based on marginal features such as the event-free survival (EFS) probability 
\begin{align}
\label{EFS}
S(t|X) = P(T > t | X) = P(Z(t)=0 | X) = \exp \biggl( - \int_{0}^{t} [ \lambda_{01}(u|X)+\lambda_{02}(u|X)] du  \biggr) \; , 
\end{align}
or the overall survival (OS) probability 
\begin{align}
\label{OS}
 P(T_2 > t | X) = P(Z(t) \in \{0,1 \} | X) = S(t|X) + \int_{0}^{t} S(u^- | X) \lambda_{01}(u | X)P_{11}(u, t | X)du \; ,  
\end{align} 
where $T_k=\inf\{t>0: Z(t)=k\}$ is the entry time to state $k$ and $T=\inf \{t>0: Z(t) \neq 0\}$ is the exit time from state $0$, $k=1,2$. Over the last two decades there has been increased interest in models based on cumulative incidence functions (CIFs) for the non-fatal event given by 
\begin{align}
\label{CIF}
F_1(t|X) = P(T_1 \leq t | X) = \int_{0}^{t} S(u^-|X) \lambda_{01}(u|X) du\; . 
\end{align}
More generally state occupancy probabilities can be considered where $P(Z(t)=k|X)=P(Z(t)=k|Z(0)=0, X)=P_{0k}(0,t|X)$ and $P_{kl}(s,t|X)=P(Z(t)=l | Z(s)=k, X), s<t$. \noindent
When models are specified for the intensity functions, (\ref{EFS}), (\ref{OS}) and (\ref{CIF}) depend on the process intensities in complex ways and there is no resulting one-dimensional summary of the treatment effect. Generalized linear models (GLMs) are often proposed to address this by modelling covariate effects directly on a marginal feature of interest. 
For the CIF in (\ref{CIF}) a GLM may take the form 
\begin{align}
\label{sec2-GM} g(F_1(t | X)) &= \alpha(t) + \beta X \; , 
\end{align}
where $g$ is a specified differentiable monotonic function on $(0,1)$, $\alpha(t)=g(F_1(t | X=0))$ is a monotonic function with $\alpha(t) \downarrow -\infty$ as $t \downarrow 0$, and $\beta=g(F_1(t | X=1))-g(F_1(t | X=0))$ is a one-dimensional marginal estimand \citep{Fine1999, Scheike2008, Gerds2012}. The function $\alpha(t)$ can be modelled parametrically \citep{Jeong2006, Jeong2007} or nonparametrically, with the latter often preferred because of perceived robustness. Estimation of $\alpha(t)$ and $\beta$ in (\ref{sec2-GM}) typically requires inverse probability of censoring weighting techniques, as we discuss in Section \ref{sec3.1}.  \newline
A Cox model for OS involves a scalar estimand $\zeta$ and has the form: 
\begin{align}
\label{sec2-Cox} P(T_2 > t| X) &= P(T_2 > t|X=0)^{\exp(\zeta X)} \; ,  
\end{align}
which is equivalent to a GLM for $P(T_2 \leq t |X)$ with $g(u)=\log(-\log(1-u)).$ A similar model is also often used for the EFS time $T$ although these are not generally compatible. 

\noindent
Despite their appeal as the basis for causal inferences in randomized trials, marginal models such as (\ref{sec2-GM}) and (\ref{sec2-Cox}) have limitations. By their nature models based on marginal features of processes provide an incomplete characterization of the process and may depend on several process intensities in a complicated way. The CIF model (\ref{sec2-GM}) does not distinguish between individuals who are alive and event-free and those who died event-free, whereas (\ref{CIF}) shows exactly how $F_1(t|X)$ depends on the process intensities.  As we illustrate in Section \ref{sec5}, secondary intensity-based analyses should accompany analyses based on a marginal feature in order to understand factors producing an observed marginal treatment effect. The adequacy of the CIF model used should also be checked. 

\section{Cumulative incidence function regression under misspecification} \label{sec3} 
Here we consider properties of estimators defined by fitting GLMs for $F_1(t|X)$ when the true data-generating process is governed by proportional intensity functions. Since such a model does not consider death following the non-fatal event of interest, it is sufficient to consider the competing risks process in Figure \ref{multistate-diagram-1}(b) with state space $\mathcal{S}=\{0,1,2\}$ and intensities $\lambda_{0k}(t|X), k=1,2$. In what follows we let $Z(T) \in \{1,2\}$ record the event type, and refer to $k=1$ as the event of interest and $k=2$ as the competing event (e.g. event-free death). Common choices for the link function $g(\cdot)$ in (\ref{sec2-GM}) include the \text{log}, \text{logit} and the \text{cloglog} link \citep{Gerds2012}. While the most suitable link function will depend on the setting, \citet{Buehler2022} observed that the \text{cloglog} model often provides a reasonable approximation in a broad range of applications; we restrict our attention to semiparametric GLMs of the form
\begin{align}
\label{EventTime}
\log \bigl(-\log(1-F_1(t | X))\bigr)=\alpha(t) + \beta X \; ,   
\end{align}
with $\alpha(t)$ left unspecified and $\beta$ defined as in Section \ref{sec-2.2}. \newline 
In Section \ref{sec3.1}, we consider estimation for (\ref{EventTime}) as described by \citet{Fine1999} (FG) and based on direct binomial (DB) regression \citep{Scheike2008}. 
Through large sample theory we derive the limiting values of the associated treatment effect estimators when (\ref{EventTime}) is violated. We consider in particular the case where the true model is a competing risks process with multiplicative treatment effects on the intensities, and study the probability limits $\beta^{\star}$ of the FG and DB estimators of $\beta$ in (\ref{EventTime}) under such a true process. These limits may also be sensitive to misspecification of the censoring model, so we comment on this first.

\subsection{Estimation methods and large sample results}
\label{sec3.1}
We let $\tau$ be the planned duration of follow-up for each individual $i=1,2,\ldots,n$, i.e. the administrative censoring time. Individuals may be prematurely lost to follow-up at a random censoring time $C_{ri} < \tau$, giving the net censoring time $C_{i}=\min(C_{ri}, \tau)$. We assume that a set of covariates $W_{i}$ is available such that $\{Z_{i}(u), u>0 \} \indep C_{ri} | X_{i}, W_{i}$ and let $G_{i}(u)=G(u|X_{i}, W_{i})=P(C_{ri}>u|X_{i},W_{i})$. If the censoring process does not depend on covariates associated with the failure process, the Kaplan-Meier (KM) estimate of $G_{i}(u)=G(u)=P(C_r>u)$ can be used for estimation of the censoring distribution. In settings where censoring depends on covariates, intensity-based regression models (e.g. Cox proportional hazards model or \citet{Aalen1980}'s additive model) are common for the censoring time $C_{r}$. If censoring depends only on a limited number of discrete covariates, an alternative to regression modeling is to define strata and use KM estimation within strata. We review estimating functions and large sample theory for KM estimation and Cox regression in Sections \ref{cov-indep-cens} and \ref{cov-dep-cens} of Appendix \ref{appendix-censoring}; in particular we review the limiting behaviour of the KM and Cox estimators of the censoring survivor function when the censoring model for $C_r$ may be misspecified. In investigations that follow we label the probability limit of the estimator as $G_{i}^{\star}(u)$ (i.e. $\widehat{G} \longrightarrow G^{\star}$), with $G_i^{\star}(u)$ representing $G^{\star}(u)$ in the case of an unstratifed KM estimator $\widehat{G}(u)$ or $G^{\star}(u|X_i,W_i)$ if a stratified KM or Cox estimator $\widehat{G}(u|X_i,W_i)$ is considered. We note that $G^{\star}=G$ when the ``working'' censoring model is correct (i.e. $\widehat{G} \longrightarrow G$). A special case is when no additional covariates $W_i$ are needed; we consider this case here and assume that $\{Z_{i}(u), u>0 \} \indep C_{ri}  \ | \ X_{i}$ for the remainder of the manuscript.

\subsubsection{Fine-Gray method of estimation}
\label{eqs-Fine-Gray} The \citet{Fine1999} method is based on the fact that (\ref{EventTime}) may be characterized as a proportional hazards model for $T_1$ with baseline cumulative subdistribution hazard function $\Gamma(t)=\exp(\alpha(t))$. Then if $Y_{i}^{\dagger}(t)=\mathbbm{1}(T_{1i} \geq t)=\mathbbm{1}(T_{i} \geq t) + \mathbbm{1}(T_{2i} \leq t)=\mathbbm{1}(Z_{i}(t^{-}) \in \{0,2\})$ is the so-called FG at-risk indicator, a weighted Cox partial likelihood leads to estimating functions
\begin{align}
\label{equationFG1}
\sum_{i=1}^{n} w_{i}(t) Y_{i}^{\dagger}(t) \bigl(&dN_{1i}(t) - \exp(\beta X_{i})d\Gamma(t) \bigr) = 0 \; ,  \\
\label{equationFG2}
\sum_{i=1}^{n} \int_{0}^{\infty} w_{i}(t) Y_{i}^{\dagger}(t) \bigl(&dN_{1i}(t) - \exp(\beta X_{i})d\Gamma(t) \bigr)X_{i} = 0 \; ,
\end{align}
for $d\Gamma(t)$ and $\beta$ respectively, where $N_{1i}(t)=\mathbbm{1}(T_{1i} \leq t)$, $\Delta N_{1i}(t) =N_{1i}(t + \Delta t^-) - N_{1i}(t^-)$ and $dN_{1i}(t)=\lim \limits_{\Delta t \downarrow 0} \Delta N_{1i}(t)$.
Note that individuals who have previously experienced the competing event (e.g. event-free death) and are thus in state $2$ at time $t^-$ are not truly at-risk of the event of interest, but the FG at-risk indicator treats them as such \citep{Putter2020}. Inverse probability of censoring weights in (\ref{equationFG1}) and (\ref{equationFG2}), defined as
\begin{align}
\label{FG-weights}
w_{i}(t)=\dfrac{\mathbbm{1}(C_i > \min(T_i,t))}{G_{i}(\min(T_i, t))} = \mathbbm{1}(t \leq \tau) \cdot \dfrac{\mathbbm{1}(C_{ri} > \min(T_i,t))}{G_{i}(\min(T_i, t))} \; , 
\end{align}
are necessary to deal with the fact that for individuals censored before $\min(T_i,t)$ we do not know the value of $N_{1i}(t)$; see Section 4.1.2 of \citet{Cook2018}. Solving (\ref{equationFG1}) for $d\Gamma(t)$ with fixed $\beta$ gives the profile estimate
\begin{equation}
\label{equationFG3}
d\widetilde{\Gamma}(t) = \sum_{i=1}^{n} w_{i}(t)Y_{i}^{\dagger}(t)dN_{1i}(t) / \sum_{i=1}^{n} w_{i}(t)Y_{i}^{\dagger}(t) \exp(\beta X_{i}) \; , 
\end{equation}
and substituting (\ref{equationFG3}) into (\ref{equationFG2}) gives the partial pseudo-score function for $\beta$ as 
\begin{equation}
\label{FGscore}
U^{FG}(\beta, G) =  \sum_{i=1}^{n} \int_{0}^{\infty} w_{i}(t) Y_{i}^{\dagger}(t) \biggl[ X_{i} - \dfrac{S^{(1)}(t, \beta)}{S^{(0)}(t, \beta)} \biggl] dN_{1i}(t) \; ,
\end{equation}
where $S^{(l)}(t, \beta) = \sum_{i=1}^{n} w_{i}(t) Y_{i}^{\dagger}(t) X_{i}^{l}\exp(\beta X_{i}), l=0,1$. Setting (\ref{FGscore}) equal to zero and solving gives the FG estimator $\widehat{\beta}_{FG}$ following estimation of $G$. The Breslow-type estimate of $\Gamma(t)=\int_{0}^{t} d\Gamma(u)$ is $\widehat{\Gamma}_{FG}(t) = \int_{0}^{t} d\widehat{\Gamma}_{FG}(u)$ where $d\widehat{\Gamma}_{FG}(u)$ can be obtained by substituting $\widehat{\beta}_{FG}$ and $\widehat{w}_{i}(t)$ into (\ref{equationFG3}):
\begin{align}
\label{equationFG3-Gammahat}
d\widehat{\Gamma}_{FG}(u) =  \sum_{i=1}^{n} \widehat{w}_{i}(u)Y_{i}^{\dagger}(u)dN_{1i}(u) / \sum_{i=1}^{n} \widehat{w}_{i}(u)Y_{i}^{\dagger}(u) \exp(\widehat{\beta}_{FG} X_{i}) \; . 
\end{align}
Three assumptions labelled A1-A3 are used in developing asymptotic results which follow. If
\begin{flalign*}
&\text{A1} \ \ \ \ \ \  \{Z(u), u>0\} \indep C_r \ | \ X \; ,  \\
&\text{A2} \ \ \ \ \ \ \text{model} \ (\ref{EventTime}) \ \text{for} \ F_1(t|X) \ \text{is valid , and} \;   \\
&\text{A3} \ \ \ \ \ \ \text{the censoring model for} \  C_r \ \text{is correctly specified such that} \  G^{\star}=G \ \text{and} \ \widehat{G} \longrightarrow G \; , &
\end{flalign*}
then $\widehat{\beta}_{FG}$ is consistent for $\beta$ and $\widehat{\Gamma}_{FG}(t)$ for $\Gamma(t)$ \citep{Fine1999, He2016}.
More generally, if $G^{\star}_{i}(\cdot)$ is defined as above and
$$ w_{i}^{\star}(t)=\mathbbm{1}(t \leq \tau) \cdot \dfrac{\mathbbm{1}(C_{ri} > \min(T_i,t))}{G_{i}^{\star}(\min(T_i, t))} \; , $$ 
then we rewrite (\ref{FGscore}) as
\begin{align}
\label{FGscore-censLimitingVals}
U^{FG}(\beta, G^{\star}) = \sum_{i=1}^{n} \int_{0}^{\infty} w_{i}^{\star}(t) Y_{i}^{\dagger}(t) \biggl[ X_{i} - \dfrac{S^{(1, \star)}(t, \beta)}{S^{(0, \star)}(t, \beta)} \biggl] dN_{1i}(t) \; , 
\end{align}
where $S^{(l, \star)}(t, \beta)=\sum_{i=1}^{n} w_{i}^{\star}(t) Y_{i}^{\dagger}(t) X_{i}^{l}\exp(\beta X_{i}), l=0,1$. The implied estimand is the probability limit $\beta^{\star}$ of the estimator $\widehat{\beta}$ based on (\ref{FGscore-censLimitingVals}), which solves 
\begin{align}
\label{ch3-limitingFG}
\mathbb{E}\bigl( U^{FG}(\beta, G^{\star}) \bigr) &= \int_{0}^{\infty} \biggl\{ s^{(1,\star)}(t) - \dfrac{s^{(1, \star)}(t, \beta)}{s^{(0,\star)}(t, \beta)} s^{(0,\star)}(t) \biggr\}dt = 0 \; , 
\end{align}
where $s^{(l,\star)}(t, \beta) = \mathbb{E}(S^{(l, \star)}(t, \beta))$ and $s^{(l, \star)}(t)=\mathbb{E} \bigl(w_{i}^{\star}(t) Y_{i}^{\dagger}(t) X_{i}^{l} dN_{1i}(t) \bigr), l=0,1$, and where expectations are taken with respect to the true competing risks, censoring and covariate processes \citep{White1982, Struthers1986}. If assumptions A1 and A3 hold but A2 does not, the expectations needed to compute (\ref{ch3-limitingFG}) are
\begin{align*}
s^{(0,\star)}(t, \beta) = s^{(0)}(t, \beta) &= \mathbb{E}_{X}\bigl((1-F_1(t|X))\exp(\beta X)\bigr) = \sum_{x=0,1} P(X=x)\exp(\beta x)(1-F_1(t|X=x)) \; ,  \\
s^{(1,\star)}(t, \beta) = s^{(1)}(t, \beta) &= \mathbb{E}_{X}\bigl((1-F_1(t|X)) X\exp(\beta X)\bigr) = P(X=1)\exp(\beta)(1-F_1(t|X=1)) \; , \\
\end{align*}
\begin{align*}
s^{(0,\star)}(t) = s^{(0)}(t) &= \mathbb{E}_{X}\bigl(\lambda_{01}(t|X)S(t|X)\bigr) =  \sum_{x=0,1} P(X=x)\lambda_{01}(t|X=x)S(t|X=x) \; , \\
s^{(1,\star)}(t) = s^{(1)}(t)&= \mathbb{E}_{X}\bigl(\lambda_{01}(t|X)S(t|X)X\bigr)= P(X=1)\lambda_{01}(t | X=1)S(t | X=1) \; , 
\end{align*}
where $s^{(l)}(t,\beta)=\mathbb{E}(S^{(l)}(t, \beta))$ and $s^{(l)}(t)=\mathbb{E} \bigl( \sum_{i=1}^{n} w_{i}(t) Y_{i}^{\dagger}(t) X_{i}^{l} dN_{1i}(t) \bigr), l=0,1$, respectively. We denote the probability limit of the solution to (\ref{ch3-limitingFG}) by $\beta^{\star}_{FG}$. Since $\mathbb{E}_{C_r|(T,Z(T)),X}(w^{\star}_{i}(t)) = 1$ under A1 and A3, $\beta^{\star}_{FG}$ is not dependent on the censoring process. The probability limit of $d\widetilde{\Gamma}(t)$ is the solution when the expected value of the LHS of (\ref{equationFG1}) is set equal to zero with fixed $\beta^{\star}_{FG}$, and can be found to be  
\begin{align*}
d\Gamma_{FG}^{\star}(t) = \dfrac{\mathbb{E}_{X}(\lambda_{01}(t|X)S(t|X))}{\mathbb{E}_{X}(\exp(\beta^{\star}_{FG} X)(1-F_1(t|X)))} = \dfrac{\sum_{x=0,1} P(X=x)\lambda_{01}(t|X=x)S(t|X=x)}{\sum_{x=0,1} P(X=x)\exp(\beta^{\star}_{FG} x)(1-F_1(t|X=x))} \; ,
\end{align*}
where $F_1(t|X)$ is given by (\ref{CIF}). Consequently, the limit of the FG estimator of $F_1(t|X)$ under (\ref{EventTime}) is 
\begin{align}
\label{F1star-FG}
F^{\star}_{1,FG}(t | X) = 1-\exp \bigl(-\exp(\alpha^{\star}_{FG}(t))\exp(\beta^{\star}_{FG}X) \bigr) \; , 
\end{align}
with $\exp(\alpha^{\star}_{FG}(t))=\Gamma^{\star}_{FG}(t)=\int_{0}^{t} d\Gamma^{\star}_{FG}(u)$.

\subsubsection{Estimating functions for direct binomial regression}
\label{sec-DB}
The binomial estimation procedure of \citet{Scheike2008} uses the fact that, at any given time $t$, $N_{1i}(t)$ follows a binomial distribution with mean $\mathbb{E}(N_{1i}(t)|X_{i})=P(T_{1i} \leq t |X_{i})=F_1(t|X_{i})$. We consider estimation of $F_1(\cdot|X_i)$ based on (\ref{EventTime}) at $R$ distinct time points $s_1, s_2, \ldots, s_R$ over $(0,\tau)$. In this case the cumulative incidence function $F_1(t|X_{i}; \theta)$ under (\ref{EventTime}) is specified at times $s_1, \ldots, s_R$ by an $(R+1)\times1$ parameter vector $\theta=(\alpha', \beta)'$, where $\alpha=(\alpha_1,\ldots,\alpha_R)'$ with $\alpha_r:=\alpha(s_r)$, $r=1,2,\ldots,R$. Estimation of $\theta$ is based on the indicators $N_{1i}(s_r)$ which are ideally available for all $r=1,2,\ldots,R$ and all $i=1,2,\ldots,n$. Note that the binary indicators $N_{11}(s_1),\ldots,N_{11}(s_R)$ from a given individual are not independent, but in the absence of random censoring one could use a composite binomial log-likelihood under a working independence assumption of the form 
\begin{align}
\label{pseudo-loglik-binomial}
\sum_{i=1}^{n} \sum_{r=1}^{R} \bigl( N_{1i}(s_r) \log(h(\alpha_r+\beta X_i)) + (1-N_{1i}(s_r))\log(1-h(\alpha_r+\beta X_i)) \bigr) \; ,
\end{align}
where $h(u)=g^{-1}(u)=1-\exp(-\exp(u))$ is the inverse of the cloglog link function. Differentiating (\ref{pseudo-loglik-binomial}) with respect to $\theta$ gives unbiased estimating functions for $\alpha$ and $\beta$. In trials with random loss to follow-up $N_{1i}(s_r)$ is known if $\min(T_i, C_i) \geq s_r$ or $T_i < \min(s_r, C_i)$, but unknown if $C_{i} < \min(T_i, s_r)$. The weighted response
\begin{align}
\label{DB-weighted-response}
\widetilde{N}_{1i}(s_r)= \dfrac{\mathbbm{1}(C_i \geq \min(T_i, s_r))N_{1i}(s_r)}{G_{i}(\min(T_i, s_r))} \overset{s_r \leq \tau}{=} \dfrac{\mathbbm{1}(C_{ri} \geq \min(T_i, s_r))N_{1i}(s_r)}{G_{i}(\min(T_i, s_r))} \;
\end{align}
is however observable when $G_{i}(\cdot)$ is known or estimated, and has the same expectation as $N_{1i}(s_r)$ if assumptions A1 and A3 are satisfied. Replacing $N_{1i}(s_r)$ by $\widetilde{N}_{1i}(s_r)$ in (\ref{pseudo-loglik-binomial}) and taking the derivatives with respect to $\theta$ gives the DB estimating equations as
\begin{align}
\label{ch3-DB-eqs}
U^{DB}(\theta, G) = \sum_{i=1}^{n} 
\begin{pmatrix}
    U_{i}(\alpha, G)  \\
    U_{i}(\beta, G) \\
\end{pmatrix} 
= \sum_{i=1}^{n}
\begin{pmatrix}
    U_{i}(\alpha_1, G)  \\
    ... \\
    U_{i}(\alpha_R, G) \\
    U_{i}(\beta, G) \\
\end{pmatrix}
= \sum_{i=1}^{n} D_{i} A^{-1}_{i}(\widetilde{N}_{1i} - \mu_i) \; , 
\end{align}
where $A_{i}=\text{diag} \bigl( \{ F_1(s_r |X_{i}; \theta)(1-F_{1}(s_r|X_{i}; \theta)) \}, r=1,\ldots,R \bigr)$ is an $R\times R$ diagonal (working independence) covariance matrix, $\widetilde{N}_{1i}=(\widetilde{N}_{1i}(s_1), \ldots, \widetilde{N}_{1i}(s_R))'$ is an $R\times1$ vector of weighted responses, $\mu_i = (\mu_i(s_1), \ldots, \mu_i(s_R))'$ is an $R\times 1$ vector with entries $\mu_i(s_r)=\mathbbm{E}(N_{1i}(s_r) | X_i; \theta) = h(\alpha_r + \beta X_{i})$, and $D_{i}=(\partial\mu_i / \partial \alpha', \partial\mu_i / \partial \beta )'$ is $(R+1) \times R$ matrix of derivatives. Following specification of the grid $s_1,\ldots,s_R$ and estimation of $G$, the DB estimator $\widehat{\theta}=(\widehat{\alpha}_{DB}', \widehat{\beta}_{DB})'$ solves $U^{DB}(\theta, G)=(0,0,\ldots,0)'$. It is possible to base estimation on all observed type 1 event times as in the FG approach, but we follow \citet{Klein2005} who recommend $5$-$10$ time points equally-spaced over $(0,\tau)$. If assumptions A1-A3 are valid, then $\widehat{\theta}$ is consistent for $\theta$ \citep{Scheike2008}.
More generally $\widehat{\theta}$ is consistent for $\theta^{\star} = ((\alpha^{\star}_{DB})', \beta_{DB}^{\star})$, the solution to $\mathbb{E}(U^{DB}(\theta, G^{\star}))=(0,0,\ldots,0)'$ where 
\begin{align}
U^{DB}(\theta, G^{\star}) = \sum_{i=1}^{n} 
\begin{pmatrix}
    U_{i}(\alpha, G^{\star})  \\
    U_{i}(\beta, G^{\star}) \\
\end{pmatrix} 
= \sum_{i=1}^{n}
\begin{pmatrix}
    U_{i}(\alpha_1, G^{\star})  \\
    ... \\
    U_{i}(\alpha_R, G^{\star}) \\
    U_{i}(\beta, G^{\star}) \\
\end{pmatrix}
= \sum_{i=1}^{n} D_{i} A^{-1}_{i}(\widetilde{N}_{1i}^{\star} - \mu_i) \;  
\end{align}
for $\widetilde{N}^{\star}_{1i}=(\widetilde{N}^{\star}_{1i}(s_1), \ldots, \widetilde{N}^{\star}_{1i}(s_R))'$ and 
$$\widetilde{N}^{\star}_{1i}(s_r)=\dfrac{\mathbbm{1}(C_{ri} \geq \min(T_i, s_r))N_{1i}(s_r)}{G^{\star}_{i}(\min(T_i, s_r))} \; , $$ 
and where expectations are taken with respect to the true competing risks, censoring and covariate processes.
The expectations needed to derive $\theta^{\star}$ under assumptions A1 and A3 are 
\begin{align}
\label{ch3-limitingDB-1}
\mathbb{E}(U_{1}(\alpha_r, G^{\star})) = \mathbb{E}(U_{1}(\alpha_r, G)) &=  \sum_{x=0,1} \dfrac{P(X=x) h'(\alpha_r+\beta x)}{h(\alpha_r+\beta x)(1-h(\alpha_r+\beta x))} (F_1(s_r | X=x) - h(\alpha_r + \beta x))  \\ 
\label{ch3-limitingDB-2}
\mathbb{E}(U_{1}(\beta, G^{\star})) = \mathbb{E}(U_{1}(\beta, G)) &= \sum_{r=1}^{R} \dfrac{P(X=1) h'(\alpha_r+\beta)}{h(\alpha_r+\beta)(1-h(\alpha_r+\beta))} (F_1(s_r | X=1) - h(\alpha_r + \beta))\; ,
\end{align}
where $F_1(s_r|X)$ is given as in (\ref{CIF}). 
The probability limit of the DB estimator of $F_1(s_r|X)$ at time $s_r$ is 
\begin{align}
\label{F1star-DB}
F^{\star}_{1,DB}(s_r|X)= h(\alpha^{\star}_{rDB} + \beta_{DB}^{\star} X) \; , \ \ \ \ \ \ \ \ \ \ \ \ \ \ r=1,2,\ldots,R.
\end{align}
\noindent 
An alternative to weighting the state $1$ occupancy indicators is to weight the binomial estimating equations (EEs) as follows:
\begin{align*}
\sum_{i=1}^{n} W_{i} D_{i} A^{-1}_{i}(N_{1i} - \mu_i) = 0 \; , 
\end{align*}
where $W_{i}$ is a $R\times R$ diagonal matrix with entries $w_{i}(s_r)$ given by (\ref{FG-weights}). Note that the limiting values are the same for both strategies when the censoring model is correctly specified. Unlike the weighted outcome approach, which uses information from all individuals at time $s_r$, individuals who were censored prior to $\min(T,s_r)$ do not contribute to the pseudo-score function at $s_r$ in the weighted EE approach. For the case of the logit model (\ref{sec2-GM}) and a single time point $s_1$ ($R=1$), \citet{Blanche2023} showed that the relative efficiency of these two strategies depends on the distribution of the censoring times. In investigations that follow we focus on the original DB method as described by \citet{Scheike2008}.

\subsection{Limiting values of FG and DB estimators under intensity-based processes} 
\label{sec-numericalstudies-CR}
\begin{figure}
    \centering
    \scalebox{0.65}{ 
    \includegraphics{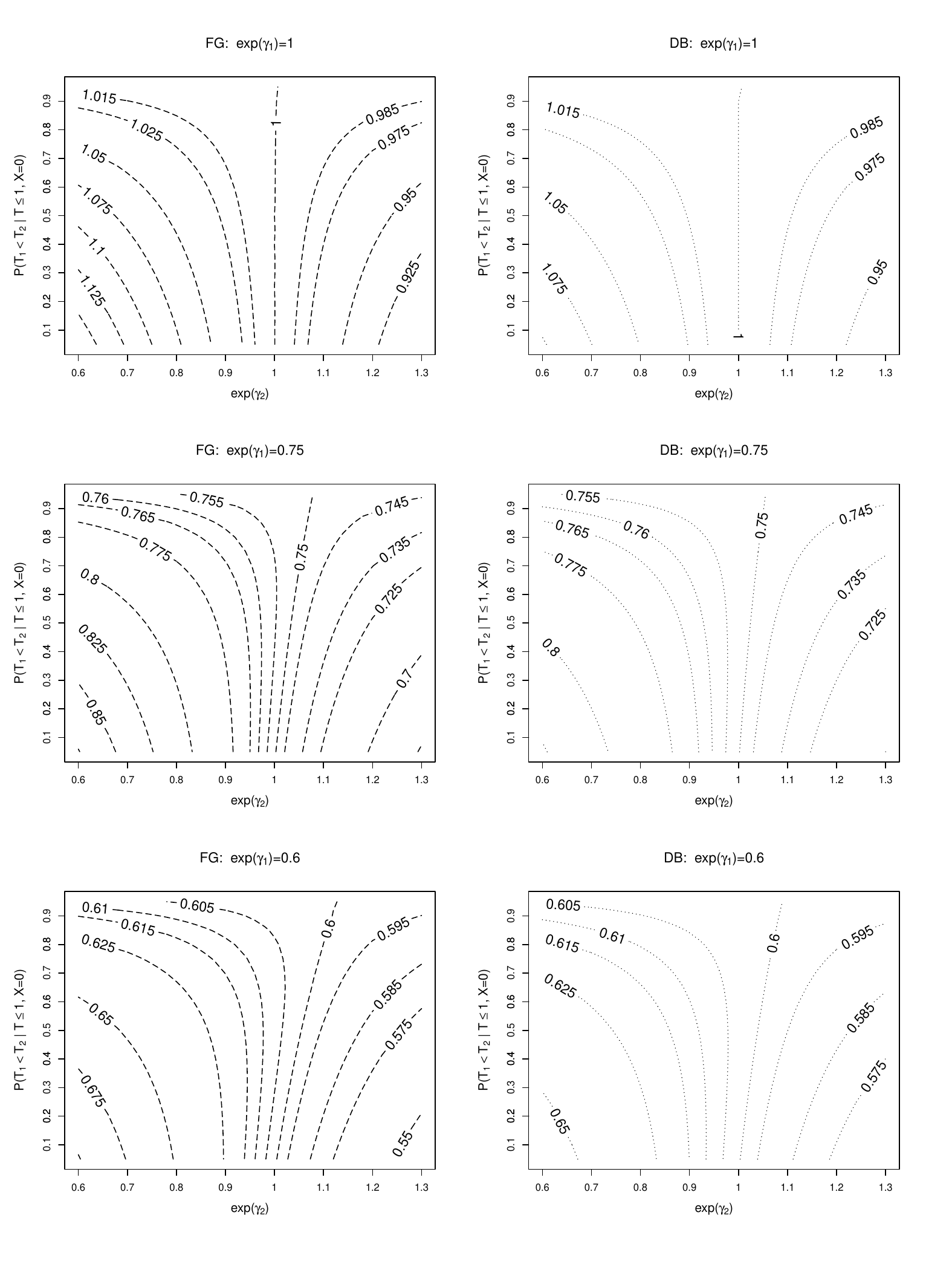}
    }
    \caption{Contours showing the limiting values $\exp(\beta^{\star}_{FG})$ (left column) and $\exp(\beta^{\star}_{DB})$ (right column) as a function of $\exp(\gamma_2)$ and $P(T_1 < T_2 | T \leq 1, X=0)$ when the true process has intensities $\lambda_{0k}(t|X)=\lambda_k\exp(\gamma_k X)$ for $\exp(\gamma_1)=1$ (top row), $0.75$ (middle row) and $0.6$ (bottom row); $\tau=1, P(T \leq 1 |X=0)=0.6$; DB estimation is based on $R=6$ equi-spaced time points in $(0,\tau), G^{\star}=G$.}
    \label{beta-star-FG-DB}
\end{figure}

Here we examine the limiting values of the FG and DB estimators using the large sample results of Section \ref{sec3.1} in a setting where the true underlying process has $0-k$ intensities of the form $\lambda_{0k}(t|X)=\lambda_{k}\exp(\gamma_k X), k=1,2$. We present corresponding results for time-inhomogeneous intensities in Section S1.2. 
Without loss of generality we set the administrative censoring time to $\tau=1$ and determined $\lambda_k$ so that $P(T \leq 1 | X = 0)=0.6$ where $T=\min(T_1, T_2)$ and $P(T_1 < T_2 | T \leq 1, X=0) \in (0,1)$. We set $\exp(\gamma_1)$ to $1,0.75$ or $0.6$ to correspond to a $0\%, 25\%$ or $40\%$ reduction in the $0-1$ intensity with treatment; values for $\exp(\gamma_2)$ varied from $0.6$ to $1.3$, giving either a decrease, no change or an increase in the $0-2$ intensity. For each parameter setting the limiting values $\beta^{\star}_{FG}$ and $\beta^{\star}_{DB}$ were computed using (\ref{ch3-limitingFG}) and (\ref{ch3-limitingDB-1})-(\ref{ch3-limitingDB-2}), respectively. \newline 
Figure \ref{beta-star-FG-DB} contains contour plots depicting the dependence of $\exp(\beta^{\star}_{FG})$ (left panels) and $\exp(\beta^{\star}_{DB})$ (right panels) on treatment and the process intensities for the case where the adopted censoring model is correct (i.e. $G^{\star}=G, \widehat{G} \longrightarrow G$) and $R=6$ for the DB approach. We do not expect $\beta^{\star}_{FG}$ and $\beta^{\star}_{DB}$ to be identical as they solve different estimating functions. If treatment does not affect the $0-1$ intensity or the $0-2$ intensity ($\exp(\gamma_1)=\exp(\gamma_2)=1$), then $\exp(\beta_{FG}^{\star})=\exp(\beta_{{DB}}^{\star})=1$; see vertical contour in upper panels. In the middle and lower panels treatment affects the $0-1$ intensity (i.e. $\exp(\gamma_1)\neq1$); if $\exp(\gamma_2)=1$, then $\exp(\beta^{\star}_{FG})$ and $\exp(\beta^{\star}_{DB})$ are very close to $\exp(\gamma_1)$. It is readily apparent from all panels of Figure \ref{beta-star-FG-DB} that treatment effects on the $0-2$ intensity (i.e. $\exp(\gamma_2)\neq1$) impact the CIF-based estimands through a reduction ($\exp(\gamma_2)>1$) or increase ($\exp(\gamma_2)<1$) in the cumulative probability of the event of interest for the experimental treatment; see also plots of $F_{1,FG}^{\star}(t|X)$ and $F^{\star}_{1,DB}(s_r|X)$ in Figures S3 and S4 of the Online Supplementary Material. The magnitude of the limiting value further depends on the probability observed events are of the cause of interest in the control arm. Naturally the greater the probability of the competing event (e.g. event-free death) the greater the impact of the treatment effect on the estimand of interest.
\newline 
We conclude that model (\ref{EventTime}) can be useful for the evaluation of experimental interventions in trials aiming to prevent type 1 events when the competing (type 2) event rate is low and treatment has a fairly small effect on the corresponding $0-2$ intensity; in this case $\beta^{\star}$ represents the effect of treatment on $F_1(t|X)$ well, and is close
to the treatment effect $\gamma_1$ on the main event intensity. More generally, $\beta^{\star}$ is a function of the true intensity-based process and the degree of model misspecification. The estimand $\beta^{\star}_{DB}$ appears less sensitive than $\beta^{\star}_{FG}$ to misspecification of (\ref{EventTime}), as $\beta^{\star}_{DB}$ is closer to $\gamma_1$ than $\beta^{\star}_{FG}$ in all scenarios considered; see also Figures S1 and S2 in Section S1.1 of the Online Supplementary Material for additional illustration. The difference between $\exp(\beta^{\star}_{FG})$ and $\exp(\beta^{\star}_{DB})$ decreases with increasing probability observed events are of type 1 in the control group. In Section S1.1 of the Online Supplementary Material we further discuss the impact of model misspecification on the estimation of $F_1(t|X)$ by comparing the true $F_1(t|X)$ (as given in (\ref{CIF})) to $F_{1,\cdot}^{\star}(t|X)$ (as given in (\ref{F1star-FG}) and (\ref{F1star-DB})) for each $X=0,1$. 
\newline  
We next provide guidance on methods for checking the assumption of a constant treatment effect in (\ref{EventTime}). We note that the cloglog form in (\ref{EventTime}) may also be inadequate in some cases. This can be examined by plotting nonparametric estimates of $F_1(t|X)$ for $X=0,1$ or by fitting a parametric family of transformation models. We illustrate the former approach in Section \ref{sec5}.

\subsection{Goodness of fit}
\label{sec3.3}
A range of methods have been suggested in the literature to assess the plausibility of a time-constant treatment effect within the FG model \citep{Zhou2013, Li2015,Boher2017}. A simple approach is to consider extended cloglog models for $F_1(t|X)$ of the form
\begin{align}
\label{gof-FG}
\log \bigl(-\log(1-F_1(t|X))\bigr)=\alpha(t)+ \beta(t)X = \alpha(t)+ \bigl[ \beta+\nu b(t) \bigr] X \; , 
\end{align}
with the time-varying treatment effect defined as $\beta(t)=\beta+\nu b(t)$ for a pre-specified function $b(t)$ of time and an unknown regression coefficient $\nu$. A Wald test of $H_{0}: \nu =0$ is thus a test of a time-constant treatment effect and can be easily executed using standard FG model software (e.g. R function crr()). Different choices of $b(t)$ are sensitive to different departures from the null hypothesis; common choices include $b(t)=t$ and $b(t)=\log(t)$. Goodness-of-fit tests based on scaled Schoenfeld residuals are common practice to assess the proportional hazards assumption in Cox regression \citep{Grambsch1994}.
\citet{Zhou2013} proposed a similar procedure based on modified weighted Schoenfeld residuals for FG-based CIF regression; in particular, they also assume $\beta(t)$ to be of the specific form $\beta+\nu b(t)$ and consider a score test of $H_0: \nu=0$ based on constrained optimization of the pseudo-score function under model (\ref{gof-FG}). Simulation studies in \citet{Zhou2013} have shown that the score test tends to be more powerful than the Wald test when the assumed form of $\beta(t)$ is correctly specified, except for the case where $\beta=0$. Alternatively, \citet{Li2015} proposed a procedure based on cumulative sums of martingale residuals to test against a general alternative of $\beta(t)$. 
\newline \noindent
Less work has been done for the DB approach. \citet{Scheike2008a} proposed nonparametric goodness-of-fit tests for $H_{0}: \beta_r=\beta$ based on Kolmogorov-Smirnov-type and Cramer-von-Mises-type statistics. Analogous to (\ref{gof-FG}), a Wald test of $H_{0}:\nu=0$ in the extended model 
\begin{align}
\label{gof-DB}
\log \bigl(-\log(1-F_1(s_r|X))\bigr)=\alpha_r + \beta_r X=\alpha_r+\bigl[\beta+\nu b(s_r)\bigr]X
\end{align}
can be considered. An alternative procedure is to carry out a Wald test of $H_{0}: \beta_r=\beta$ for all $r=1,2,\ldots,R$, or equivalently $H_{0}: C\beta^{\dagger}=0$ for $\beta^{\dagger}=(\beta_1,\ldots,\beta_R)'$, using the test statistic
\begin{align}
\label{wald-DB}
\bigl( C\widehat{\beta}^{\dagger} \bigr)' \bigl(C \widehat{\Sigma}_{\beta}^{\dagger}C' \bigl )^{-1} C\widehat{\beta}^{\dagger} \ \sim \ \chi^{2}_{R-1} \; ,  
\end{align}
and with $\widehat{\beta}^{\dagger}=(\widehat{\beta}_1, \ldots,\widehat{\beta}_R)'$ assumed to be approximately normal with mean $\beta^{\dagger}$ and covariance matrix $\Sigma_\beta^{\dagger}$. The so-called $(R-1)\times R$ contrast matrix $C$ has non-zero entries $c_{jl}=1$ if $j=l$ and $c_{jl}=-1$ if $l=j+1, j=1,\ldots,R-1, l=1,\ldots,R$. We refer to Section 4.1.3 of \citet{Cook2018} for details on estimation of $\theta^{\dagger}=(\alpha_1,\ldots, \alpha_R, {\beta^{\dagger}}')'$ in model (\ref{gof-DB}). Bootstrap methods as outlined in Section 11.4.2 of \citet{Klein2014} can be used to obtain $\widehat{\Sigma}_{\beta}^{\dagger}$. 

\section{Testing treatment effects in cumulative incidence function regression under misspecification}
\label{sec-testing}
Hypothesis testing is central to the evaluation of experimental treatments in randomized trials, so we consider this topic with two aims. We first discuss a variety of null hypotheses that may be tested with competing risks processes, examine their relationships, and provide a short overview of existing tests. Second, using large sample theory and empirical studies we investigate the behaviour of Wald tests of no treatment effect on $F_1(t|X)$ in settings where (\ref{EventTime}) is misspecified. 

\subsection{Formulation of testing framework and robustness property for FG-/DB-based Wald tests}
\label{sec-4.1}
\begin{figure}[ht]
\centering
\adjustbox{scale=1.1, center}{
\begin{tikzcd}[arrows=Rightarrow]
& H_{0}^{S} & H_{0}^{F_1F_2} \arrow[ddl] \arrow[dl] 
\arrow[dddl, shift left=1.1ex] \arrow[l, shift right=1.3ex] \\ 
& H_{0}^{\lambda_1} \\
H_{0}^{\lambda_1\lambda_2} \arrow[uurr, Leftrightarrow, shift right=-0.1ex, bend left] \arrow[uur,bend left] \arrow[r] \arrow[ur, shift right=-0.2ex] \arrow[dr, shift right=0.5ex] & H_{0} & 
\\
& H_{0}^{\lambda_2} 
\end{tikzcd}}
\caption{Relationships between different null hypotheses in a competing risks process.}
\label{fig-null-hypos}
\end{figure}
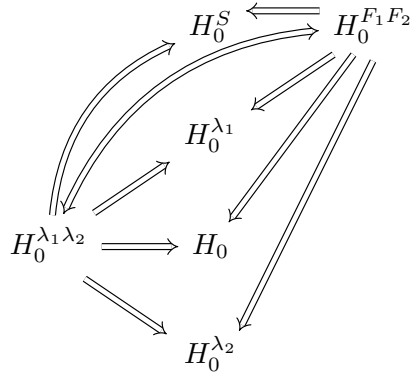

Several different null hypotheses can be formulated for testing treatment effects within the context of Figure \ref{multistate-diagram-1}(b), including 
\begin{align}
\label{H0-global} &H_{0}^{\lambda_1\lambda_2}: \ \lambda_{0k}(t|X=1) = \lambda_{0k}(t|X=0) \ \text{for} \ t \geq 0 \ \text{and all} \ k=1,2 \\
\label{H0-local} &H_{0}^{\lambda_k}: \ \lambda_{0k}(t|X=1) = \lambda_{0k}(t|X=0) \ \text{for} \ t \geq 0 \\
\label{H0} &H_{0}: \ F_1(t|X=1) = F_1(t|X=0) \ \text{for} \ t \geq 0  \\
\label{H0-S} &H_{0}^{S}: \ S(t|X=1) = S(t|X=0) \ \text{for} \ t \geq 0 \\    
\label{H0-F1F2} &H_{0}^{F_1F_2}: \ F_k(t|X=1) = F_k(t|X=0) \ \text{for} \ t \geq 0 \ \text{and all} \ k=1,2 \; , 
\end{align}
and with the cumulative incidence function for competing events in (\ref{H0-F1F2}) defined as $$F_2(t|X)=P(T_2 \leq t|X)= \int_{0}^{t}  S(u^-|X)\lambda_{02}(u|X)du \; . $$
A schematic showing the relation between the null hypotheses (\ref{H0-global})-(\ref{H0-F1F2}) is given in Figure \ref{fig-null-hypos}. The global null $H_{0}^{\lambda_1 \lambda_2} = H_{0}^{\lambda_1} \cap H_{0}^{\lambda_2}$ corresponds to no treatment effect on the full disease process and, thus, no effect on any process feature. The null hypothesis $H_{0}^{\lambda_k}$ is weaker than $H_{0}^{\lambda_1\lambda_2}$ as it reflects no treatment effect only on the $0-k$ intensity, $k=1,2$. Neither $H_{0}^{\lambda_1}$ nor $H_{0}^{\lambda_2}$ alone imply $H_{0}^{S}$ or $H_{0}$ due to (\ref{EFS}) and (\ref{CIF}). For example, if $H_{0}^{\lambda_1}$ is true but $H_{0}^{\lambda_2}$ is not, then $F_1(t|X)$ depends on $X$. Clearly, $H_{0}^{\lambda_k}$ alone does not imply $H_{0}^{F_1F_2}$ either. The null hypothesis $H_{0}$ is aligned with the estimator of $\beta$ such that if $\beta=0$ in (\ref{EventTime}) then $F_1(t|X)=F_1(t)$. Mathematically, however, $H_0$ does not imply $H_{0}^{\lambda_1}, H_{0}^{\lambda_2}$ or $H_{0}^{\lambda_1 \lambda_2}$, but it does impose complicated constraints on the two intensities. For example, $f_1(t|X=1)=f_1(t|X=0)$ implies that
$$\lambda_{01}(t|X=0)e^{-(\Lambda_{01}(t|X=0) + \Lambda_{02}(t|X=0))} = \lambda_{01}(t|X=1)e^{-(\Lambda_{01}(t|X=1) + \Lambda_{02}(t|X=1))} \; . $$
Such a constraint would not be satisfied with most intensity-based models. Since $S(t|X)=1-F_1(t|X)-F_2(t|X)$ and $\lambda_{0k}(t|X)=f_k(t|X)/S(t|X)$, $H_{0}^{F_1F_2}$ implies $H_{0}^{S}, H_{0}^{\lambda_k}$ and $H_0^{\lambda_1 \lambda_2}$.
If $H_{0}$ is true but $H_{0}^{F_2}: F_2(t|X=1)=F_2(t|X=0)$ is not, then $S(t|X)$ depends on $X$. Finally, $H_{0}^{S}$ does not imply any of the other null hypotheses in (\ref{H0-global})-(\ref{H0}) and (\ref{H0-F1F2}).

\noindent
A key question concerning tests of hypotheses (\ref{H0-global})-(\ref{H0-F1F2}) is: if a test has nominal size $\omega$, what is the actual (empirical) size $\omega^{\star}$ under a different null hypothesis? For instance, if a test of $H_{0}$ is based on a marginal model such as (\ref{EventTime}), what is the rejection rate if the true process satisfies $H_{0}^{\lambda_1}$? We discuss specific tests in detail shortly, but first state the following theorem pertaining to Figure \ref{fig-null-hypos}.

\begin{theorem} \label{theorem1} Consider tests $T^{\lambda_1\lambda_2}, T^{\lambda_k}, T^{F_1}, T^{S}$ and $T^{F_1F_2}$ of nominal size $\omega$ for the null hypotheses (\ref{H0-global})-(\ref{H0-F1F2}), respectively. 
Under $H_{0}^{\lambda_1\lambda_2}$, $P(T^{\lambda_k}$ rejects $ H_{0}^{\lambda_k})=\omega, P(T^{F_1}$ rejects $H_{0})=\omega, P(T^{S}$ rejects $H_0^{S})=\omega$ and $P(T^{F_1F_2}$ rejects $H_0^{F_1F_2})=\omega$. \newline
Under $H_{0}^{\lambda_k}$, $P(T^{\lambda_1\lambda_2}$ rejects $H_{0}^{\lambda_1\lambda_2}) \neq \omega, P(T^{F_1}$ rejects $H_{0}) \neq \omega, P(T^{S}$ rejects $H_0^{S})\neq \omega$ and $P(T^{F_1F_2}$ rejects $H_0^{F_1F_2})\neq \omega$, with the actual rejection rate in all cases dependent on $\lambda_{02}(t|X)$ if $k=1$ and $\lambda_{01}(t|X)$ if $k=2$. \newline
Under $H_{0}$, $P(T^{\lambda_1\lambda_2}$ rejects $H_{0}^{\lambda_1\lambda_2}) \neq \omega, P(T^{\lambda_k}$ rejects $H_{0}^{\lambda_k}) \neq \omega, P(T^{S}$ rejects $H_0^{S})\neq \omega$ and $P(T^{F_1F_2}$ rejects $H_0^{F_1F_2})\neq \omega$, with the actual rejection rate in all cases dependent on $\lambda_{01}(t|X)$ and $\lambda_{02}(t|X)$. 
\newline 
Under $H_{0}^{S}$, $P(T^{\lambda_1\lambda_2}$ rejects $ H_{0}^{\lambda_1\lambda_2}) \neq \omega, P(T^{\lambda_k}$ rejects $H_{0}^{\lambda_k}) \neq \omega, P(T^{F_1}$ rejects $H_0)\neq \omega$ and $P(T^{F_1F_2}$ rejects $H_0^{F_1F_2}) \neq \omega$, with the actual rejection rate in all cases dependent on both $\lambda_{01}(t|X)$ and $\lambda_{02}(t|X)$. \newline 
Under $H_{0}^{F_1F_2}, P(T^{\lambda_1\lambda_2}$ rejects $H_0^{\lambda_1\lambda_2})=\omega,  P(T^{\lambda_k}$ rejects $ H_{0}^{\lambda_k})=\omega, P(T^{F_1}$ rejects $H_{0})=\omega$ and $P(T^{S}$ rejects $H_0^{S})=\omega$.
\end{theorem}

\noindent
The local null $H_{0}^{\lambda_1}$ is typically tested using the nonparametric log-rank test for the event of interest \citep{Mantel1966,Peto1972,Cox1972}, which is asymptotically equivalent to a Wald (or score) test for $\gamma_1=0$ based on the Cox model $\lambda_{01}(t|X)=\lambda_1(t)\exp(\gamma_1 X)$. In the simulations that follow we refer to these two test statistics as $T^{\lambda_1}_{LR}$ and $T^{\lambda_1}_{Cox}$. Similar tests can be considered for $H_{0}^{\lambda_2}$. A common test for the global null $H_{0}^{\lambda_1\lambda_2}$ is based on a Wald statistic for the hypothesis $\gamma_1=\gamma_2=0$ in the joint Cox model $\lambda_{01}(t|X)=\lambda_1(t)\exp(\gamma_1 X), \lambda_{02}(t|X)=\lambda_2(t)\exp(\gamma_2 X)$. Since $\bigl[\widehat{\gamma}_1/se(\widehat{\gamma}_1)\bigr]^2 \sim \chi^2_{1}$ and $\bigl[\widehat{\gamma}_2/se(\widehat{\gamma}_2)\bigr]^2 \sim \chi^2_{1}$ are asymptotically chi-squared and independent, this is a two degrees of freedom test and we label the test statistic as $T^{\lambda_1\lambda_2}_{Cox}$. In a similar way, one could construct a $\chi^{2}$ joint test based on two independent log-rank test statistics \citep{Li2016}. The null $H_{0}^{S}$ can be tested based on a log-rank test statistic for the event-free survival time $T$, or based on a Wald statistic for $\zeta=0$ in a cloglog model of the form (\ref{sec2-Cox}); we disregard tests of $H_{0}^{S}$ in our simulations. Two common tests for $H_{0}$ are a generalized log-rank test due to \citet{Gray1988} and a Wald test based on (\ref{EventTime}). We label the Gray test statistic as $T^{F_1}_{Gray}$ and the Wald tests based on the FG and DB estimation procedures for $\beta$ as $T^{F_1}_{FG}$ and $T^{F_1}_{DB}$, respectively. \newline 
Tests for $H_{0}^{F_1F_2}$ have not received much attention. \citet{Shi2013} proposed a Wald test for the hypothesis $\theta=0$ based on parametric models of the form $F_1(t|X)=1-(1-F_1(\infty|X=0)v(t; b_1, c_1))^{\exp(\theta X)}$ and $F_2(t|X)=(1-F_1(\infty|X=0))^{\exp(\theta X)}v(t; b_2, c_2)$, where $F_1(t|X=0)=F_1(\infty|X=0)v(t; b_1,c_1)$ satisfies a modified three-parameter logistic function with $v(t; b_k,c_k) \longrightarrow 1$ as $t \longrightarrow \infty$, $b_k>0, c_k \in \mathbb{R}$. We discuss limitations of this class of models in Section \ref{sec4.2-setup}. Alternatively, a Wald statistic for $\theta_1=\theta_2=0$ in a multinomial logistic model of the form $\log(F_k(t|X)/S(t|X))=A_k(t)+\theta_k X, k=1,2$ could be considered \citep{Gerds2012}.
Various other less commonly used tests have been discussed in the literature \citep{Li2016, Dobler2017, Tiwari2006}; we do not consider them here. 
\newline \newline \noindent
For the remainder of this section we focus on the FG- and DB-based Wald tests $T_{FG}^{F_1}$ and $T_{DB}^{F_1}$ for $\beta=0$ and assess their properties under model misspecification. We let $\widehat{\beta}$ be the solution to the FG pseudo-score function (\ref{FGscore}) or the DB estimating equations (\ref{ch3-DB-eqs}), respectively. Statistics of the form $(\widehat{\beta} - 0)/se(\widehat{\beta})$ have in general an asymptotic $N(0,1)$ distribution under $\beta=0$, provided the underlying modeling assumptions are all satisfied. With a misspecified model (\ref{EventTime}) the null hypothesis actually being tested is $\beta^{\star}_{FG}=0$ or $\beta^{\star}_{DB}=0$, respectively, based on 
\begin{align}
\label{large-sample-FG-DB}
\sqrt{n}(\widehat{\beta}_{FG}-\beta^{\star}_{FG}) \xrightarrow{\mathcal{D}} N(0, V_{FG}) \ \ \ \text{or} \ \ \
\sqrt{n}(\widehat{\beta}_{DB}-\beta^{\star}_{DB}) \xrightarrow{\mathcal{D}} N(0, V_{DB}) \; , 
\end{align}
where the robust sandwich-type variance under the FG procedure takes the form $V_{FG}= \Omega(\beta^{\star}_{FG})/I^2(\beta^{\star}_{FG})$ for $I(\beta)=\mathbbm{E}\bigl(-\partial U^{FG}(\beta, G^{\star})/\partial \beta\bigr)$ and $\Omega(\beta)=\mathbb{E}\bigl(\bigl[U^{FG}(\beta, G^{\star})\bigr]^2\bigr) + \Psi$ \citep{Fine1999, He2016}.
The second term of $\Omega(\beta)$ reflects the uncertainty in estimating the censoring distribution $G$; an explicit expression for $\Psi$ can be found in Section $4$ in \citet{Fine1999} for the case of covariate-independent censoring and in Appendix A of \citet{He2016} for covariate-dependent censoring. The asymptotic variance $V_{DB}=\widetilde{\Omega}(\beta^{\star}_{DB})/\tilde{I}^2(\beta^{\star}_{DB})$ under the DB procedure is likewise of robust sandwich-type form; see Eq. (7) in \citet{Scheike2008} for further details. We note that robust variance estimation for $\widehat{\beta}$ is available for FG and DB regression in the R functions crr(), comp.risk() and cifreg() \citep{timereg-package, Holst2016}. \newline
\noindent
It is important to understand under which circumstances a Wald test based on $(\widehat{\beta}-0)/se(\widehat{\beta})$ remains asymptotically valid even if (\ref{EventTime}) is misspecified. We conclude this section by stating our main result in the following theorem; a proof of Theorem \ref{theorem2} is provided in Appendix \ref{app-proof-theorem2}. 

\begin{theorem}
\label{theorem2}
When the \textit{cloglog} model (\ref{EventTime}) for $F_1(t|X)$ is misspecified, a hypothesis test based on the Wald statistic $(\widehat{\beta}-0)/se(\widehat{\beta})$ with a robust standard error estimate is asymptotically valid if the probability limit $\beta^{\star}$ of $\widehat{\beta}$ is $0$ under the true process. As before, let $\widehat{\Gamma}_{FG}(t)$ denote the Breslow-type estimator for $\Gamma(t)=\exp(\alpha(t))$ under the FG procedure, and let $\widehat{\alpha}_r$ be the estimator for $\alpha_r=g(F_1(s_r|X=0))$ under DB estimation, $r=1,2,\ldots,R$. Assume $P(X=1)=0.5$ and the regularity conditions in Appendix A.1 of \citet{He2016} and Appendix A of \citet{Scheike2008} such that $d\widehat{\Gamma}_{FG}(t) \longrightarrow d\Gamma^{\star}_{FG}(t),\widehat{\beta}_{FG} \longrightarrow \beta_{FG}^{\star}$ and $\widehat{\alpha}_{rDB} \longrightarrow \alpha^{\star}_{rDB}, \widehat{\beta}_{DB} \longrightarrow \beta_{DB}^{\star}$.
\vspace*{-2.3mm}
\begin{enumerate}
\item If the FG procedure in Section \ref{eqs-Fine-Gray} is used for estimation with a correctly specified censoring model ($G^{\star}=G$), then $\beta_{FG}^{\star}=0$ under $H_{0}^{\lambda_1\lambda_2}$ and $H_{0}$, but $\beta_{FG}^{\star}\neq 0$ under $H_{0}^{\lambda_1}$ alone.  
\item If we consider DB estimation at $R$ distinct time points $s_1,s_2,\ldots,s_R$ (as in Section \ref{sec-DB}) with a correctly specified censoring model, and treat the baseline parameters $\alpha_r$ as unconstrained, then $\beta_{DB}^{\star}=0$ under $H_{0}^{\lambda_1\lambda_2}$ and $H_{0}$, but $\beta_{DB}^{\star}\neq 0$ under $H_{0}^{\lambda_1}$. 
\end{enumerate}
\vspace*{-2.3mm}
To summarize, the FG- and DB-based Wald tests of $\beta=0$ are robust when the true process satisfies $H_{0}^{\lambda_1 \lambda_2}$ or $H_{0}$, in the sense that they control the nominal level $\omega$ in large samples. This robustness property does not hold if robust variance estimates are not used. 
\end{theorem} 

\subsection{Some numerical results for testing null hypotheses}
\label{simulations-testing}
We report on simulation studies designed to assess the type I error rates of the FG- and DB-based Wald tests and their power for various alternatives. Although the main focus here is on tests based on cumulative incidence function regression models, we also examined the empirical rejection rates for a selection of intensity-based tests including $T_{Cox}^{\lambda_1\lambda_2}, T_{LR}^{\lambda_1}$ and $T_{Cox}^{\lambda_1}$ in order to confirm our findings in Theorem \ref{theorem1} in empirical studies. \cite{Freidlin2005} and \cite{Williamson2007} consider testing with competing risks processes based on the framework of latent failure times. The assumptions of this framework are impossible to assess in any given setting, unlike the intensity-based models we consider here. 

\subsubsection{Simulation setup}
\label{sec4.2-setup}
The binary treatment indicator $X$ was simulated from a binomial distribution with $P(X=1)=0.5$. To simulate competing risks data under $H_{0}^{\lambda_1\lambda_2}$ and $H_{0}^{\lambda_1}$ we generated processes with intensity functions $\lambda_{0k}(t|X)=\lambda_{k}\exp(\gamma_k X)$ following the procedure in Section 2.5 of \citet{Cook2018}; for each of $X=0,1$ the event time $T | X$ was first generated from an exponential distribution with rate $\lambda_{01}(t|X)+\lambda_{02}(t|X)$. Given $T=t$, the event type was then drawn from a Bernoulli trial with probability $\lambda_{01}(t|X)/(\lambda_{01}(t|X)+\lambda_{02}(t|X))$ that the event was of the cause of interest. As in Section \ref{sec-numericalstudies-CR}, with $\tau=1$ we set the values for the time-homogeneous baseline intensities $\lambda_{1}$ and $\lambda_{2}$ to satisfy the constraints: $P(T \leq 1 | X = 0)=0.6$ and $P(T_1 < T_2 | T \leq 1, X=0)=0.4, 0.6$ and $0.8$. This gave type 1 event rates by $\tau=1$ for individuals in the control group of $F_1(1|X=0)=0.24, 0.36$ and $0.48$ and type 2 event rates of $F_2(1|X=0)=0.36,0.24$ and $0.12$, respectively. We further set $\exp(\gamma_1)$ to $0.6, 0.75, 0.9$ and $1$, and $\exp(\gamma_2)$ to $0.5, 0.8, 0.9, 1, 1.1$ and $1.5$. Note that $H_{0}^{\lambda_1 \lambda_2}$ is true when $\exp(\gamma_1)=\exp(\gamma_2)=1$ and $H_{0}^{\lambda_1}$ is true when $\exp(\gamma_1)=1$. The random withdrawal time $C_r$ was simulated from an exponential distribution with rate $\rho>0$ set so that $\pi_{r}=P(C_r < T_1 | T_1 \leq \min(T_2, 1))=\mathbb{E}_{X}(P(C_r < T_1 | T_1 \leq \min(T_2, 1), X))$ is $0.2$, so $20\%$ of the type 1 events occurring before the administrative censoring time are right-censored due to random loss to follow-up. 
\newline 
Section 5.3.6 of \citet{Beyersmann2009} discusses methods for simulating competing risks processes where $F_1(t|X)$ has the cloglog form.
With their indirect simulation method, models for $F_1(t|X)$ and $F_2(t|X)$ take the form 
\vspace*{-1mm}
\begin{align}
\nonumber g(F_1(t|X)) &= g\bigl(F_1(\infty|X=0)(1-\exp[-t]) \bigr) + \beta X \\ &\Longleftrightarrow \ \  
\label{simH0-model1-JB} F_1(t|X) = 1- \bigl(1-F_1(\infty|X=0)(1-\exp[-t])\bigr)^{\exp(\beta X)}\; \\
\label{simH0-model2-JB} \log(F_2(t|X=1)) &=  \exp(\beta)\log(F_2(\infty|X=0)) + B(t; \beta) \; , 
\end{align}
where $g(u)=\text{cloglog}(u), q:=F_1(\infty|X=0)$ is the proportion of type 1 events at $t=\infty$ in the control group and $F_2(\infty|X)=1-F_1(\infty|X)$, $X=0,1$. Limitations to this approach include the fact that (\ref{simH0-model2-JB}) does not have a standard GLM form since $B(t; \beta)$ is a function of $\beta$, and more importantly, $F_1(t|X)$ and $F_2(t|X)$ depend on the same $\beta$. If $\beta=0$ in (\ref{simH0-model1-JB}), then $F_1(t|X)=F_1(t)$ and $F_2(t|X)=F_2(t)$, and also $\lambda_{0k}(t|X)=\lambda_{0k}(t)$ for $k=1,2$; thus within the class of models given by (\ref{simH0-model1-JB}) and (\ref{simH0-model2-JB}) $H_{0}$ implies $H_{0}^{F_2}, H_{0}^{F_1F_2}, H_{0}^{\lambda_k}$ and $H_{0}^{\lambda_1\lambda_2}$. We therefore also considered models for $F_1(t|X)$ and $F_2(t|X)$ as follows: 
\begin{align}
\nonumber g(F_1(t|X; \beta)) &= g\bigl(F_1(\infty|X=0)(1-\exp[-\psi_1(t)]) \bigr) + \beta X \\ \label{simH0-model1-approach2} &\Longleftrightarrow \ \ F_1(t|X) = 1-\bigl(1-F_1(\infty|X=0)(1-\exp[-\psi_1(t)])\bigr)^{\exp(\beta X)} \; ,  \\ 
\label{simH0-model2-approach2} F_2(t|X; \beta, \beta_2) &= (1-F_1(\infty|X))g^{-1}(\alpha_2(t) + \beta_2 X) \; , 
\end{align}
where $F_1(t|X=0)=q(1-\exp[-\psi_1(t)]), F_1(\infty|X)=1-(1-q)^{\exp(\beta X)}$ and $\alpha_2(t)=\log(\psi_2(t))$ for $\psi_k(t)=t$, in which case (\ref{simH0-model1-approach2}) reduces to (\ref{simH0-model1-JB}). 
If $\beta=0$ in (\ref{simH0-model1-approach2}), then $F_2(\infty|X=1)=F_2(\infty|X=0)$. If $\beta_2=0$ and $\beta\neq0$, then $F_2(t|X=1) \neq F_2(t|X=0)$. These models are both too restrictive for general use, but they provide a framework to check on the properties of tests under $H_0$ and alternatives based on (\ref{EventTime}). More generally we note that if $f_k(t|X) = \partial F_k(t|X) / \partial t$ is the subdensity function of $T_k$ and $B(t; \beta)=\log\bigl(1-\exp(-t\exp(\beta))\bigr)$ as in \citet{Beyersmann2009}, we obtain the $0-k$ intensities under models (\ref{simH0-model1-JB})-(\ref{simH0-model2-JB}) and (\ref{simH0-model1-approach2})-(\ref{simH0-model2-approach2}) as
\begin{align}
\label{intensities-H0}
\lambda_{0k}^{\dagger}(t|X) &= \lim_{\Delta t \downarrow 0} \dfrac{P(T \in [t,  t+\Delta t^-), Z(T)=k \ | \ T \geq t, X)}{\Delta t} = \dfrac{f_k(t|X)}{1-F_1(t|X)-F_2(t|X)} \; .
\end{align} 
We further note that both approaches satisfy the constraint $0 \leq F_1(t|X)+F_2(t|X) \leq 1$. Given $X$ and $\lambda_{0k}^{\dagger}(t|X)$, the event time $T|X$ can then be generated using the inversion method: we first simulated a standard uniform random variable $U$, set the realized value $u$ equal to $1-S(t|X)=F_1(t|X)+F_2(t|X)$ and solved for $t$. Given $T=t$, the event type $k$ can be determined by a Bernoulli experiment which assigns it to be a type 1 event with probability $\lambda_{01}^{\dagger}(t|X)/(\lambda_{01}^{\dagger}(t|X)+\lambda_{02}^{\dagger}(t|X)), k=1,2$. With $\tau=1$, we specified $q$ so that $F_1(1|X=0)=0.24, 0.36$ and $0.48$, and set $\exp(\beta)$ to $0.8,0.9,1$ and $1.1$, and $\exp(\beta_2)$ to $0.8$ and $1$. This gave type 2 event rates by $\tau=1$ for individuals in the control group of $F_2(\tau|X=0)=0.39, 0.27$ and $0.15$ under both approaches (\ref{simH0-model1-JB})-(\ref{simH0-model2-JB}) and (\ref{simH0-model1-approach2})-(\ref{simH0-model2-approach2}). As before, the random withdrawal time $C_r$ follows an exponential distribution with rate $\rho >0$ set to satisfy $\pi_r=0.2$. \newline
For each null model and parameter setting we consider the following tests: a two-d.f. Wald test for $H_{0}^{\lambda_1\lambda_2}$ based on Cox models for $\lambda_{01}(t|X),\lambda_{02}(t|X)$ or $\lambda_{01}^{\dagger}(t|X),\lambda_{02}^{\dagger}(t|X)$ ($T^{\lambda_1\lambda_2}_{Cox}$), the log-rank test for $H_{0}^{\lambda_1}$ ($T^{\lambda_1}_{LR}$), a one-d.f. Wald test for $H_{0}^{\lambda_1}$ based on a Cox cause-specific hazards model for $\lambda_{01}(t|X)$ or $\lambda_{01}^{\dagger}(t|X)$ ($T^{\lambda_1}_{Cox}$), Gray's test for $H_{0}$ ($T^{F_1}_{Gray}$), a one-d.f. Wald test for $H_{0}$ based on the FG procedure with a robust variance estimate ($T^{F_1}_{FG}$), and a one-d.f. Wald test for $H_{0}$ based on DB regression with $R=6$ or $R=3$ equi-spaced time points on $(0,1)$ and a robust variance estimate ($T^{F_1}_{DB_R}$). We simulated datasets with $n=1000$ individuals and conducted the tests using the coxph(), survdiff(), cuminc(), crr() and comp.risk() functions in the respective R packages survival, cmprsk and timereg as mentioned before  \citep{survival-package, timereg-package}; the crr() and comp.risk() functions for the FG and DB procedures automatically provide robust variance estimates. Moreover, unstratified KM estimation was used in these functions to get an estimate of the common censoring distribution $G(u)=P(C_r>u)$. We note that if $w(t)$ in the FG procedure is replaced by stabilized weights of the form $G(t)\mathbbm{1}(C_r \geq \min(T,t))/G(\min(T,t))$, Theorem \ref{theorem2} continues to hold. We then calculated the empirical rejection rates as the proportion of $n_{sim}=10000$ simulations in which the respective null hypothesis was rejected at the nominal $5\%$ level ($\omega=0.05$): with this many simulation if the test rejects the null at the nominal $5\%$ level, we expect the empirical rejection rate to fall within the interval $[0.0457, 0.0543]$ $95\%$ of the time. 

\subsubsection{Simulation results}
\label{sim-results}
\begin{figure}[ht]
\centering
   \scalebox{0.50}{
    \includegraphics{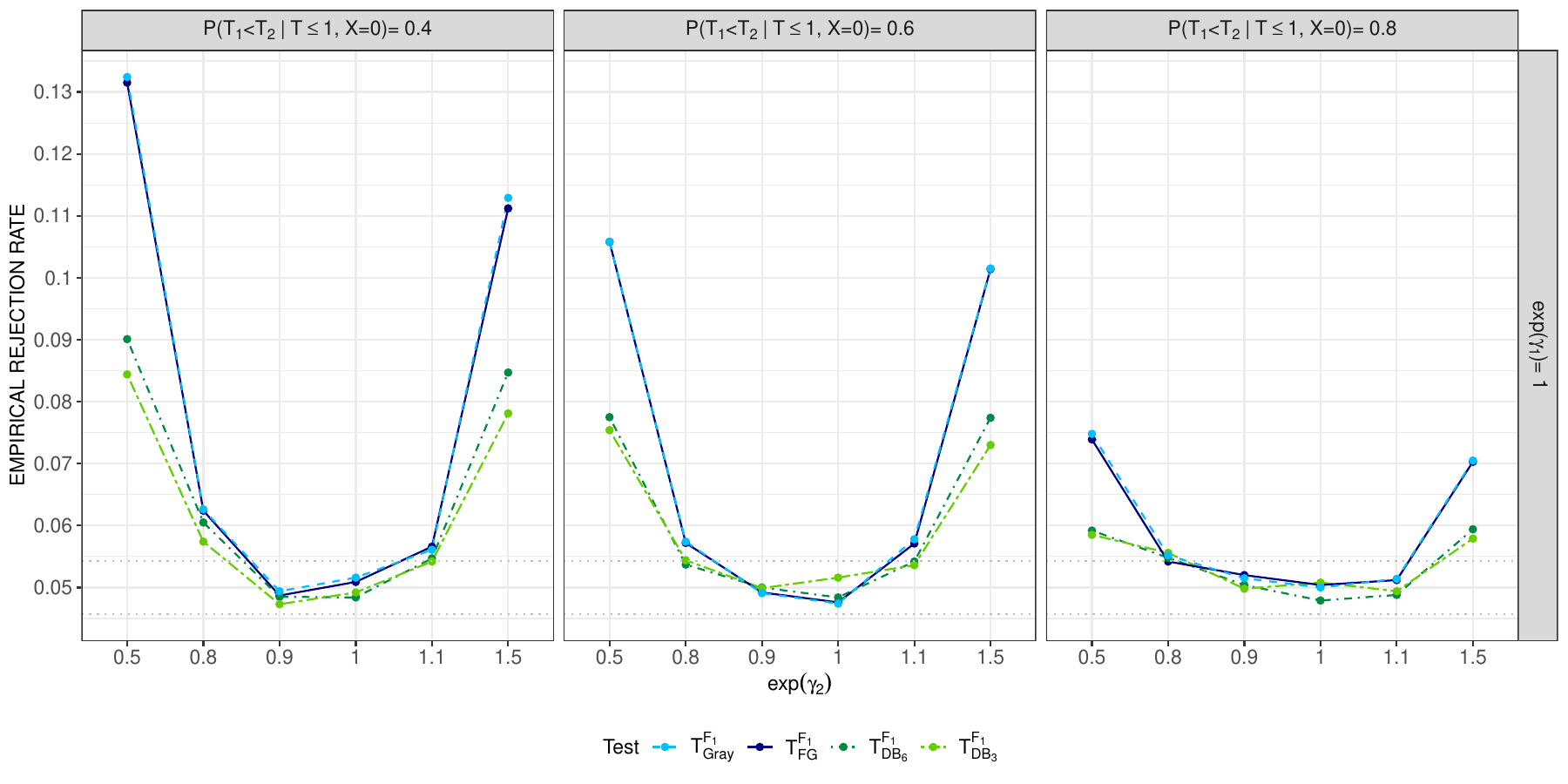}
    }
    \caption{Empirical type I error rates of tests $T^{F_1}$ under $H_{0}^{\lambda_1\lambda_2 }$ ($\exp(\gamma_1)=\exp(\gamma_2)=1$) and $H_{0}^{\lambda_1}$ ($\exp(\gamma_1)=1$) for different values of $\exp(\gamma_2)$ and $P(T_1 < T_2 | T \leq 1, X=0)=0.4$ (left panel), $0.6$ (middle panel) and $0.8$ (right panel), nominal test size is $\omega=0.05; n=1000$ individuals, $n_{sim}=10000$, $\tau=1, P(T \leq 1 | X=0)=0.6, \pi_r=0.2$.}
    \label{fig-t1e-cr}
\end{figure}

Table \ref{sim-table-t1e-power} reports the empirical rejection rates of all tests considered for the setting where the data were generated according to the multiplicative intensity-based models and where $F_1(1|X=0)=0.36$ and $F_2(1|X=0)=0.24$ (i.e. $P(T_1<T_2 | T \leq 1, X=0)=0.6$), along with the corresponding limiting values $\beta^{\star}_{FG}, \beta^{\star}_{DB_6}$ and $\beta^{\star}_{DB_3}$ under FG and DB estimation. The 13$th$ row (when $\exp(\gamma_1)=\exp(\gamma_2)=1$) shows the results under the true process implied by $H_{0}^{\lambda_1\lambda_2}$; the 1$st$ row of each block (when $\exp(\gamma_1)=1$) represents the empirical type I error rates when $H_{0}^{\lambda_1}$ is true. In addition to Table \ref{sim-table-t1e-power}, the empirical type I error rates for tests $T^{F_1}_{Gray}, T^{F_1}_{FG}, T^{F_1}_{DB_6}$ and $T^{F_1}_{DB_3}$ under $H_{0}^{\lambda_1\lambda_2}$ and $H_{0}^{\lambda_1}$ are also graphically depicted in the middle panel of Figure \ref{fig-t1e-cr}.

\begin{table}[h]
\centering
\scalebox{0.85}{ 
\begin{tabular}{cccccccccccccccc}
\multicolumn{1}{l}{\multirow{4}{*}{}} & \multicolumn{1}{l}{} & \multicolumn{1}{l}{}                  & \multicolumn{9}{c}{REJECTION RATES}                                                                                                                                                                                                                                                                                                            &                      & \multicolumn{3}{c}{}                                                                                                                              \\ \cline{4-12}
\multicolumn{1}{l}{}                  & \multicolumn{1}{l}{} & \multicolumn{1}{l}{}                  & \multicolumn{2}{c}{\multirow{2}{*}{$H_{0}^{\lambda_1}$}}                       & \multicolumn{1}{l}{} & \multicolumn{1}{l}{\multirow{2}{*}{$H_{0}^{\lambda_1\lambda_2}$}} & \multicolumn{1}{l}{} & \multicolumn{4}{c}{\multirow{2}{*}{$H_{0}$}}                                                                                                & \multicolumn{1}{l}{} & \multicolumn{1}{l}{}                                      & \multicolumn{1}{l}{}                      & \multicolumn{1}{l}{}                      \\
\multicolumn{1}{l}{}                  &                      & \multicolumn{1}{l}{}                  & \multicolumn{2}{c}{}                                                           &                      & \multicolumn{1}{l}{}                                              &                      & \multicolumn{4}{c}{}                                                                                                                        &                      & \multicolumn{3}{c}{PROBABILITY LIMITS}                                                                                                            \\ \cline{4-5} \cline{7-7} \cline{9-12} \cline{14-16} 
\multicolumn{1}{l}{}                  &                      & \multicolumn{1}{l}{}                  & \multirow{2}{*}{$T^{\lambda_1}_{LR}$} & \multirow{2}{*}{$T^{\lambda_1}_{Cox}$} &                      & \multirow{2}{*}{$T^{\lambda_1\lambda_2}_{Cox}$}                   &                      & \multirow{2}{*}{$T^{F_1}_{Gray}$} & \multirow{2}{*}{$T^{F_1}_{FG}$} & \multirow{2}{*}{$T^{F_1}_{DB_6}$} & \multirow{2}{*}{$T^{F_1}_{DB_3}$} & \multirow{6}{*}{}    & \multicolumn{1}{l}{\multirow{2}{*}{$\beta_{FG}^\star$}} & \multirow{2}{*}{$\beta_{DB_6}^\star$} & \multirow{2}{*}{$\beta_{DB_3}^\star$} \\
$\exp(\gamma_2)$                        & $\exp(\gamma_1)$       & \multicolumn{1}{l}{\multirow{5}{*}{}} &                                       &                                        &                      &                                                                   &                      &                                   &                                 &                                   &                                   &                      & \multicolumn{1}{l}{}                                      &                                           &                                           \\ \cline{1-2} \cline{4-5} \cline{7-7} \cline{9-12} \cline{14-16} 
\multirow{4}{*}{0.5}                  & 1                    & \multicolumn{1}{l}{}                  & 0.0479                                & 0.0476                                 &                      & 0.9691                                                            &                      & 0.1058                            & 0.1058                          & 0.0775                            & 0.0754                            &                      & 0.0825                                                    & 0.0593                                    & 0.0560                                    \\
                                      & 0.9                  & \multicolumn{1}{l}{}                  & 0.1466                                & 0.1459                                 &                      & 0.9771                                                            &                      & 0.0557                            & 0.0550                          & 0.0617                            & 0.0649                            &                      & -0.0210                                                   & -0.0453                                   & -0.0487                                   \\
                                      & 0.75                 & \multicolumn{1}{l}{}                  & 0.6549                                & 0.6535                                 &                      & 0.9938                                                            &                      & 0.3702                            & 0.3697                          & 0.3849                            & 0.3753                            &                      & -0.2003                                                   & -0.2264                                   & -0.2299                                   \\
                                      & 0.6                  & \multicolumn{1}{l}{}                  & 0.9791                                & 0.9786                                 &                      & 0.9997                                                            &                      & 0.9055                            & 0.9052                          & 0.8905                            & 0.8803                            &                      & -0.4205                                                   & -0.4483                                   & -0.4520                                   \\
\multicolumn{16}{l}{}                                                                                                                                                                                                                                                                                                                                                                                                                                                                                                                                                                                                            \\
\multirow{4}{*}{0.8}                  & 1                    &                                       & 0.0477                                & 0.0470                                 &                      & 0.2303                                                            &                      & 0.0574                            & 0.0572                          & 0.0537                            & 0.0544                            &                      & 0.0324                                                    & 0.0234                                    & 0.0221                                    \\
                                      & 0.9                  &                                       & 0.1350                                & 0.1337                                 &                      & 0.3093                                                            &                      & 0.0840                            & 0.0832                          & 0.0886                            & 0.0864                            &                      & -0.0706                                                   & -0.0808                                   & -0.0822                                   \\
                                      & 0.75                 &                                       & 0.6481                                & 0.6464                                 &                      & 0.6923                                                            &                      & 0.5238                            & 0.5229                          & 0.4879                            & 0.4755                            &                      & -0.2495                                                   & -0.2614                                   & -0.2630                                   \\
                                      & 0.6                  &                                       & 0.9763                                & 0.9762                                 &                      & 0.9734                                                            &                      & 0.9509                            & 0.9511                          & 0.9295                            & 0.9166                            &                      & -0.4691                                                   & -0.4828                                   & -0.4846                                   \\
\multicolumn{16}{l}{}                                                                                                                                                                                                                                                                                                                                                                                                                                                                                                                                                                                                            \\
\multirow{4}{*}{0.9}                  & 1                    &                                       & 0.0478                                & 0.0473                                 &                      & 0.0862                                                            &                      & 0.0491                            & 0.0492                          & 0.0500                            & 0.0499                            &                      & 0.0161                                                    & 0.0117                                    & 0.0110                                    \\
                                      & 0.9                  &                                       & 0.1403                                & 0.1398                                 &                      & 0.1600                                                            &                      & 0.1106                            & 0.1113                          & 0.1057                            & 0.1046                            &                      & -0.0868                                                   & -0.0924                                   & -0.0932                                   \\
                                      & 0.75                 &                                       & 0.6477                                & 0.6464                                 &                      & 0.5771                                                            &                      & 0.5824                            & 0.5804                          & 0.5289                            & 0.5120                            &                      & -0.2655                                                   & -0.2728                                   & -0.2738                                   \\
                                      & 0.6                  &                                       & 0.9740                                & 0.9737                                 &                      & 0.9548                                                            &                      & 0.9592                            & 0.9593                          & 0.9398                            & 0.9306                            &                      & -0.4850                                                   & -0.4941                                   & -0.4953                                   \\
\multicolumn{16}{l}{}                                                                                                                                                                                                                                                                                                                                                                                                                                                                                                                                                                                                            \\
\multirow{4}{*}{1}                    & 1                    &                                       & 0.0493                                & 0.0484                                 &                      & 0.0534                                                            &                      & 0.0474                            & 0.0476                          & 0.0484                            & 0.0516                            &                      & 0.0000                                                    & 0.0000                                    & 0.0000                                    \\
                                      & 0.9                  &                                       & 0.1446                                & 0.1430                                 &                      & 0.1171                                                            &                      & 0.1390                            & 0.1401                          & 0.1296                            & 0.1228                            &                      & -0.1028                                                   & -0.1040                                   & -0.1041                                   \\
                                      & 0.75                 &                                       & 0.6406                                & 0.6388                                 &                      & 0.5304                                                            &                      & 0.6211                            & 0.6219                          & 0.5610                            & 0.5402                            &                      & -0.2814                                                   & -0.2842                                   & -0.2846                                   \\
                                      & 0.6                  &                                       & 0.9729                                & 0.9727                                 &                      & 0.9469                                                            &                      & 0.9697                            & 0.9688                          & 0.9470                            & 0.9373                            &                      & -0.5007                                                   & -0.5054                                   & -0.5060                                   \\
\multicolumn{16}{l}{}                                                                                                                                                                                                                                                                                                                                                                                                                                                                                                                                                                                                            \\
\multirow{4}{*}{1.1}                  & 1                    &                                       & 0.0542                                & 0.0537                                 &                      & 0.0872                                                            &                      & 0.0578                            & 0.0571                          & 0.0542                            & 0.0536                            &                      & -0.0159                                                   & -0.0116                                   & -0.0109                                   \\
                                      & 0.9                  &                                       & 0.1426                                & 0.1419                                 &                      & 0.1540                                                            &                      & 0.1699                            & 0.1687                          & 0.1479                            & 0.1420                            &                      & -0.1187                                                   & -0.1154                                   & -0.1150                                   \\
                                      & 0.75                 &                                       & 0.6322                                & 0.6297                                 &                      & 0.5614                                                            &                      & 0.6597                            & 0.6585                          & 0.5942                            & 0.5706                            &                      & -0.2971                                                   & -0.2956                                   & -0.2954                                   \\
                                      & 0.6                  &                                       & 0.9696                                & 0.9692                                 &                      & 0.9484                                                            &                      & 0.9741                            & 0.9740                          & 0.9516                            & 0.9435                            &                      & -0.5163                                                   & -0.5166                                   & -0.5166                                   \\
\multicolumn{16}{l}{}                                                                                                                                                                                                                                                                                                                                                                                                                                                                                                                                                                                                            \\
\multirow{4}{*}{1.5}                  & 1                    &                                       & 0.0519                                & 0.0509                                 &                      & 0.7824                                                            &                      & 0.1015                            & 0.1014                          & 0.0774                            & 0.0730                            &                      & -0.0780                                                   & -0.0569                                   & -0.0540                                   \\
                                      & 0.9                  &                                       & 0.1385                                & 0.1375                                 &                      & 0.8145                                                            &                      & 0.3136                            & 0.3123                          & 0.2422                            & 0.2246                            &                      & -0.1805                                                   & -0.1604                                   & -0.1577                                   \\
                                      & 0.75                 &                                       & 0.6225                                & 0.6203                                 &                      & 0.9387                                                            &                      & 0.7994                            & 0.7989                          & 0.7107                            & 0.6787                            &                      & -0.3585                                                   & -0.3400                                   & -0.3376                                   \\
                                      & 0.6                  &                                       & 0.9701                                & 0.9699                                 &                      & 0.9955                                                            &                      & 0.9923                            & 0.9920                          & 0.9769                            & 0.9705                            &                      & -0.5774                                                   & -0.5605                                   & -0.5583                                  
\end{tabular}
}
\caption{Empirical rejection rates for different statistical tests of nominal size $\omega=0.05$ when the true process has intensities $\lambda_{0k}(t|X)=\lambda_{k}\exp(\gamma_k X)$, along with the limiting values $\beta_{FG}^\star$ and $\beta_{DB}^\star$; $\gamma_1: \log(0.9)=-0.1054, \log(0.75)=-0.2877, \log(0.6)=-0.5108$; $n=1000$ individuals, $n_{sim}=10000$, $\tau=1$, $P(T \leq 1 | X=0)=0.6, P(T_1 < T_2 | T \leq 1, X=0)=0.6, \pi_r=0.2$.}
\label{sim-table-t1e-power}
\end{table}

\newpage

Consistent with Theorem \ref{theorem2}, the FG- and DB-based Wald tests $T^{F_1}_{FG}, T^{F_1}_{DB_6}$ and $T^{F_1}_{DB_3}$ are valid under $H_{0}^{\lambda_1\lambda_2}$ in the sense that their type I error rate is compatible with the nominal $5 \%$ level. Gray's test for $H_{0}$, $T^{F_1}_{Gray}$, also shows good type I error control under $H_{0}^{\lambda_1\lambda_2}$. Under $H_{0}^{\lambda_1}$, all tests for $H_{0}$ have inflated type I error rates except when $\exp(\gamma_2) = 1$. The inflation is very small when $\exp(\gamma_2)$ is close to $1$ (see e.g. the 1$st$ row of the 3$rd$ block), but increases as the magnitude of $\exp(\gamma_2)$ increases and as the proportion of type 1 events in the control group decreases; see Figure \ref{fig-t1e-cr}. This agrees with our theoretical results presented in Theorem \ref{theorem1}. The FG statistics $T^{F_1}_{Gray}$ and $T^{F_1}_{FG}$ have a larger type I error inflation than $T^{F_1}_{DB_6}$ and $T^{F_1}_{DB_3}$; note that $\beta^{\star}_{DB_6}$ and $\beta^{\star}_{DB_3}$ are closer to $0$ than $\beta^{\star}_{FG}$.
The tests $T_{LR}^{\lambda_1}$ and $T_{Cox}^{\lambda_1}$ for $H_{0}^{\lambda_1}$ yield very similar type I error rates consistent with the nominal $5 \%$ level under $H_{0}^{\lambda_1\lambda_2}$ and $H_{0}^{\lambda_1}$, as expected. The test $T^{\lambda_1\lambda_2}_{Cox}$ maintains control over the type I error rate under $H_{0}^{\lambda_1\lambda_2}$, but has inflated type I error rates under $H_{0}^{\lambda_1}$; as per Theorem \ref{theorem1}, 
the inflation varies with the magnitude of $\exp(\gamma_2)$ and $P(T_1 < T_2 | T \leq 1, X=0)$; see Tables S1 and S2 in Section S2 of the Online Supplementary Material where $P(T_1 < T_2 | T \leq 1, X=0)=0.4$ and $0.8$. 
\newline 
Simulation results for scenarios where the data were generated according to models (\ref{simH0-model1-JB})-(\ref{simH0-model2-JB}) or (\ref{simH0-model1-approach2})-(\ref{simH0-model2-approach2}) and where $F_1(1|X=0)=0.36$ are given in Table \ref{sim-table-H0-1}. The 3$rd$ row of the first block and the 5$th$ and 6$th$ rows of the second block (when $\exp(\beta)=1$) represent the empirical rejection rates under $H_{0}$. For the former case, since $H_{0}$ implies $H_{0}^{\lambda_1}$ and $H_{0}^{\lambda_1 \lambda_2}$ within this class of models, all tests have good type I error control. For the latter case, the empirical type I error rates of tests $T_{FG}^{F_1}, T_{DB_6}^{F_1}, T_{DB_3}^{F_1}$ and $T_{Gray}^{F_1}$ are close to the nominal level, confirming the findings in Theorem \ref{theorem2}. The intensity-based tests $T_{LR}^{\lambda_1}, T_{Cox}^{\lambda_1}$ and $T_{Cox}^{\lambda_1\lambda_2}$ control the nominal level if $\exp(\beta_2)=1$, in which case (\ref{simH0-model1-approach2})-(\ref{simH0-model2-approach2}) reduces to (\ref{simH0-model1-JB})-(\ref{simH0-model2-JB}). When there is a separate effect of treatment on $F_2(t|X)$ (i.e. $\exp(\beta_2) \neq 1$), these tests are based on non-proportional intensities and hence give inflated type I error rates; see also Table S3 in Section S2.

\begin{landscape}
\begin{figure}
\centering
    \scalebox{0.70}{
    \includegraphics{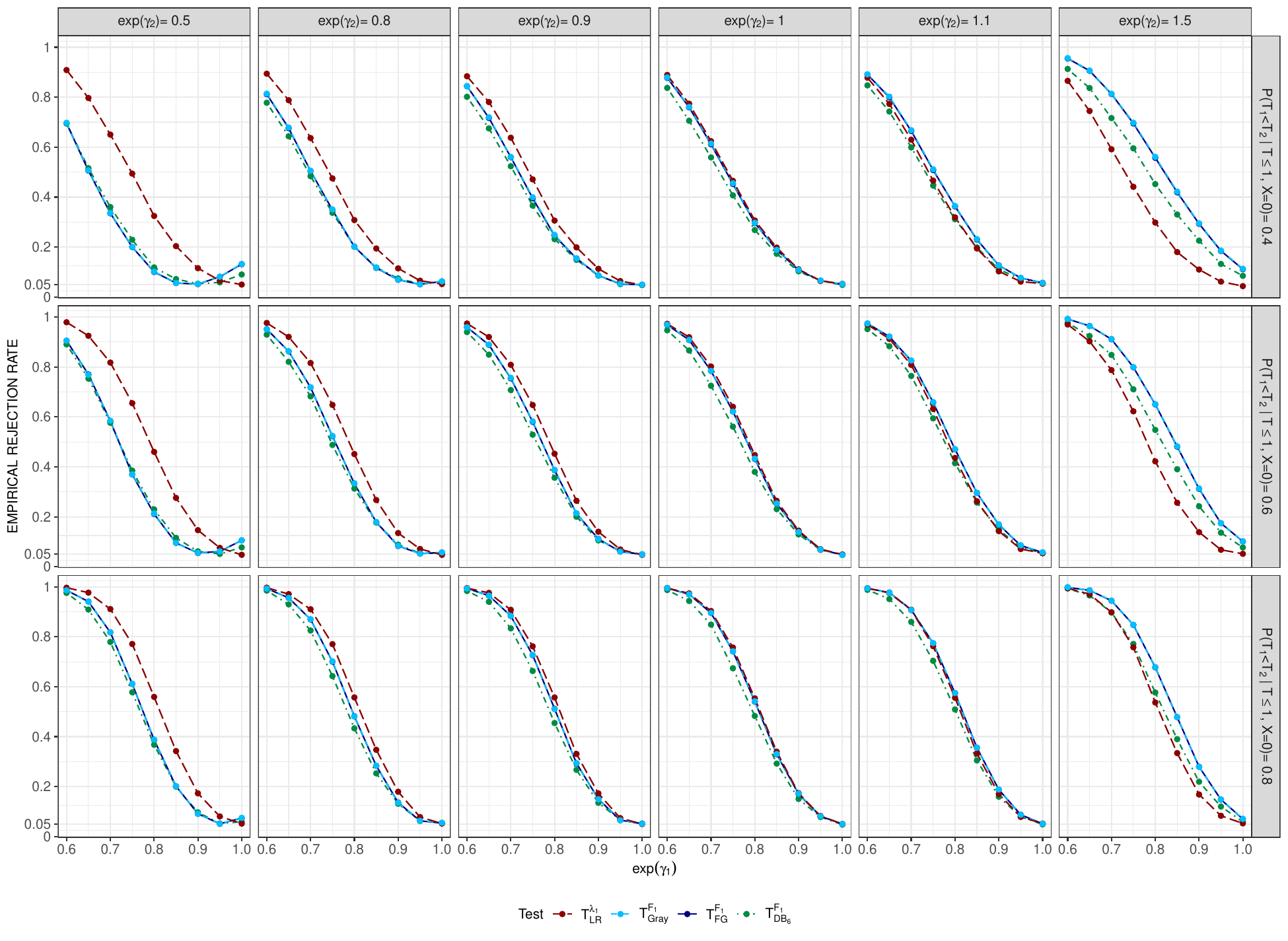}
    }
     \caption{Empirical rejection rates for different statistical tests under various alternatives; $n=1000$ individuals, $n_{sim}=10000$, $\tau=1$, $P(T \leq 1 | X=0)=0.6, P(T_1 < T_2 | T \leq 1, X=0)=0.4$ (top row), $0.6$ (middle row) and $0.8$ (bottom row), $\pi_r=0.2$.}
     \label{fig-power2-sim}
\end{figure}
\end{landscape}

\noindent
We now consider the power of the tests under a variety of different alternatives and begin with the results presented in Tables \ref{sim-table-t1e-power}, S1 and S2 where $P(T_1 < T_2 | T \leq 1, X=0)=0.6, 0.4$ and $0.8$, respectively. As one would expect, the power of all tests increases with increasing magnitude of $\gamma_1$, the treatment effect on the $0-1$ intensity, and increasing proportion of type $1$ events, here through $P(T_1 < T_2 | T \leq 1, X=0)$; see also Figure \ref{fig-power2-sim}. The power of tests for $H_{0}^{\lambda_1}$ is unaffected by $\gamma_2$, with $T^{\lambda_1}_{LR}$ and $T^{\lambda_1}_{Cox}$ being equally powerful. These results are well-known, and we now compare $T^{\lambda_1}$ and $T^{F_1}$ in terms of power. \newline
First we note that the FG statistics $T^{F_1}_{Gray}$ and $T^{F_1}_{FG}$ are equally powerful in all scenarios. The power of $T^{F_1}_{FG}, T^{F_1}_{DB_6}$ and $T^{F_1}_{DB_3}$ is in general affected by the corresponding probability limits $\beta^{\star}$ and the asymptotic variance estimators of $\widehat{\beta}$. Although $\beta_{DB_R}^\star$ is quite robust to the number of time points $R$ (i.e. $\beta_{DB_6}^\star \approx \beta_{DB_3}^\star$), the asymptotic variances become larger with decreasing $R$; the DB$_{6}$ analysis is thus expected to be more powerful than the DB$_{3}$ analysis, and this is what our simulation study reveals. Compared to $T_{FG}^{F_1}$, the DB$_6$ statistic leads to a less powerful analysis in most settings. 

\noindent
Figure \ref{fig-power2-sim} depicts the power of $T^{\lambda_1}_{LR}, T^{F_1}_{Gray}, T^{F_1}_{FG}$ and $T^{F_1}_{DB_6}$ as a function of $\exp(\gamma_1)$ and $\exp(\gamma_2)$ for the different values of $P(T_1 < T_2 | T \leq 1, X=0)$. When $\exp(\gamma_1)<1$ and $\exp(\gamma_2)=1$, the tests $T^{\lambda_1}_{LR}, T^{F_1}_{Gray}$ and $T^{F_1}_{FG}$ achieve comparable power, while $T^{F_1}_{DB_6}$ is slightly less powerful; see the third set of panels from the right. If instead $\exp(\gamma_2)<1$ (or $\exp(\gamma_2)>1$), the tests $T^{F_1}_{Gray}, T^{F_1}_{FG}$ and $T^{F_1}_{DB_6}$ are less (or more) powerful than $T^{\lambda_1}_{LR}$; see the first three sets of panels from the left and the first two sets of panels from the right, respectively. However, when $\exp(\gamma_2)>1$, the tests for $H_{0}$ have inflated type I error rates. Interestingly, when there is only a relatively small treatment effect on the $0-2$ intensity (i.e. $\exp(\gamma_2)=1.1$), $T^{F_1}_{DB_6}$ is less powerful than $T^{\lambda_1}_{LR}$. 
\newline 
Naturally as the treatment effect $\beta$ on $F_1(t|X)$ in Table \ref{sim-table-H0-1} decreases, the power of the FG- and DB-based Wald tests $T^{F_1}_{FG}, T^{F_1}_{DB_6}$ and $T^{F_1}_{DB_3}$ decreases as well. If data were generated by models (\ref{simH0-model1-approach2})-(\ref{simH0-model2-approach2}), the power of $T^{F_1}_{FG}, T^{F_1}_{DB_6}$ and $T^{F_1}_{DB_3}$ is unaffected by $\beta_2$. 

\begin{table}[h]
\centering
\scalebox{0.85}{ 
\begin{tabular}{ccclccccccccccc}
\multicolumn{3}{l}{\multirow{4}{*}{}}                                                                                                    & \multirow{17}{*}{} & \multicolumn{9}{c}{REJECTION RATES}                                                                                                                                                                                                                                                                                      & \multirow{17}{*}{} & \multicolumn{1}{l}{}                                                              \\ \cline{5-13}
\multicolumn{3}{l}{}                                                                                                                     &                    & \multicolumn{2}{c}{\multirow{2}{*}{$H_{0}^{\lambda_1}$}}                       & \multirow{16}{*}{} & \multirow{2}{*}{$H_{0}^{\lambda_1\lambda_2}$}   & \multirow{16}{*}{} & \multicolumn{4}{c}{\multirow{2}{*}{$H_{0}^{F_1}$}}                                                                                          &                    & \multirow{2}{*}{\begin{tabular}[c]{@{}c@{}}PROBABILITY \\ LIMITS\end{tabular}}    \\
\multicolumn{3}{l}{}                                                                                                                     &                    & \multicolumn{2}{c}{}                                                           &                    &                                                 &                    & \multicolumn{4}{c}{}                                                                                                                        &                    &                                                                                   \\ \cline{5-6} \cline{8-8} \cline{10-13} \cline{15-15} 
\multicolumn{3}{l}{}                                                                                                                     &                    & \multirow{2}{*}{$T^{\lambda_1}_{LR}$} & \multirow{2}{*}{$T^{\lambda_1}_{Cox}$} &                    & \multirow{2}{*}{$T^{\lambda_1\lambda_2}_{Cox}$} &                    & \multirow{2}{*}{$T^{F_1}_{Gray}$} & \multirow{2}{*}{$T^{F_1}_{FG}$} & \multirow{2}{*}{$T^{F_1}_{DB_6}$} & \multirow{2}{*}{$T^{F_1}_{DB_3}$} &                    & \multirow{2}{*}{$\beta^{\star}_{FG}, \beta^{\star}_{DB_6}, \beta^{\star}_{DB_3}$} \\
Models                                                                      & $\exp(\beta)$        & \multicolumn{1}{l}{$\exp(\beta_2)$} &                    &                                       &                                        &                    &                                                 &                    &                                   &                                 &                                   &                                   &                    &                                                                                   \\ \cline{1-3} \cline{5-6} \cline{8-8} \cline{10-13} \cline{15-15} 
\multirow{4}{*}{\begin{tabular}[c]{@{}c@{}}(4.7)\\ -\\ (4.8)\end{tabular}}  & 0.8                  & \multirow{4}{*}{-}                  &                    & 0.4921                                & 0.4906                                 &                    & 0.3975                                          &                    & 0.4530                            & 0.4514                          & 0.4022                            & 0.3851                            &                    & -0.2231                                                                           \\
                                                                            & 0.9                  &                                     &                    & 0.1569                                & 0.1554                                 &                    & 0.1208                                          &                    & 0.1441                            & 0.1447                          & 0.1327                            & 0.1279                            &                    & -0.1054                                                                           \\
                                                                            & 1                    &                                     &                    & 0.0493                                & 0.0489                                 &                    & 0.0482                                          &                    & 0.0473                            & 0.0472                          & 0.0475                            & 0.0464                            &                    & 0.0000                                                                            \\
                                                                            & 1.1                  &                                     &                    & 0.1338                                & 0.1326                                 &                    & 0.1045                                          &                    & 0.1264                            & 0.1267                          & 0.1144                            & 0.1123                            &                    & 0.0953                                                                            \\ \cline{1-3} \cline{5-6} \cline{8-8} \cline{10-13} \cline{15-15} 
\multirow{8}{*}{\begin{tabular}[c]{@{}c@{}}(4.9)\\ -\\ (4.10)\end{tabular}} & \multirow{2}{*}{0.8} & 0.8                                 &                    & 0.4820                                & 0.4791                                 &                    & 0.3865                                          &                    & 0.4441                            & 0.4433                          & 0.3926                            & 0.3737                            &                    & \multirow{2}{*}{-0.2231}                                                          \\
                                                                            &                      & 1                                   &                    & 0.3673                                & 0.3654                                 &                    & 0.4230                                          &                    & 0.4496                            & 0.4491                          & 0.3982                            & 0.3810                            &                    &                                                                                   \\ \cline{2-3} \cline{5-6} \cline{8-8} \cline{10-13} \cline{15-15} 
                                                                            & \multirow{2}{*}{0.9} & 0.8                                 &                    & 0.1866                                & 0.1850                                 &                    & 0.1969                                          &                    & 0.1432                            & 0.1423                          & 0.1279                            & 0.1229                            &                    & \multirow{2}{*}{-0.1054}                                                          \\
                                                                            &                      & 1                                   &                    & 0.1207                                & 0.1199                                 &                    & 0.1248                                          &                    & 0.1437                            & 0.1427                          & 0.1276                            & 0.1267                            &                    &                                                                                   \\ \cline{2-3} \cline{5-6} \cline{8-8} \cline{10-13} \cline{15-15} 
                                                                            & \multirow{2}{*}{1}   & 0.8                                 &                    & 0.0750                                & 0.0738                                 &                    & 0.4432                                          &                    & 0.0501                            & 0.0500                          & 0.0488                            & 0.0507                            &                    & \multirow{2}{*}{0.0000}                                                           \\
                                                                            &                      & 1                                   &                    & 0.0492                                & 0.0488                                 &                    & 0.0480                                          &                    & 0.0472                            & 0.0471                          & 0.0474                            & 0.0463                            &                    &                                                                                   \\ \cline{2-3} \cline{5-6} \cline{8-8} \cline{10-13} \cline{15-15} 
                                                                            & \multirow{2}{*}{1.1} & 0.8                                 &                    & 0.0709                                & 0.0703                                 &                    & 0.3573                                          &                    & 0.1322                            & 0.1319                          & 0.1186                            & 0.1167                            &                    & \multirow{2}{*}{0.0953}                                                           \\
                                                                            &                      & 1                                   &                    & 0.1130                                & 0.1122                                 &                    & 0.1149                                          &                    & 0.1311                            & 0.1307                          & 0.1211                            & 0.1164                            &                    &                                                                                  
\end{tabular}
}
\caption{Empirical rejection rates for different statistical tests of nominal size $\omega=0.05$ when the true process is implied by models (\ref{simH0-model1-JB})-(\ref{simH0-model2-JB}) and (\ref{simH0-model1-approach2})-(\ref{simH0-model2-approach2}),
along with the limiting values $\beta_{FG}^\star$ and $\beta_{DB}^{\star}$; $\beta:\log(0.8)=-0.2231, \log(0.9)=-0.1054, \log(1.1)=0.0953$; $n=1000$ individuals, $n_{sim}=10000$, $\tau=1, F_1(1|X=0)=0.36, F_2(1|X=0)=0.27, \pi_r=0.2$.}
\label{sim-table-H0-1} 
\end{table}

\subsubsection{Implications of simulation studies and recommendations}
Here we make some concluding remarks on the key results of our simulation studies: 
\begin{enumerate}
    \item If the primary objective of a trial is in testing $H_{0}$, we recommend use of robust Wald statistics such as $T_{FG}^{F_1}$ and $T_{DB_R}^{F_1}$, e.g. $R=6$. We have shown that these tests are asymptotically valid under $H_{0}^{\lambda_1 \lambda_2}$ and $H_{0}$ in the sense that they control the type I error. With true processes satisfying $H_{0}^{\lambda_1}$ alone however, the type I error is close to the nominal level $\omega$ only if $\beta^{\star} \approx0$; in settings where both $\exp(\gamma_2)$ and $P(T_1 < T_2 | T \geq 1, X=0)$ are not too far from $1$ the limiting values $\beta^{\star}_{FG}$ and $\beta^{\star}_{DB}$ are found to be roughly $0$. For the case where $\exp(\gamma_2)=0.8$ and $P(T_1 < T_2 | T \leq 1, X=0)=0.6$, for instance, the empirical rejection rates for $T_{FG}^{F_1}$ and $T_{DB_6}^{F_1}$ were at most $0.057$ when $n=1000$, $0.055$ when $n=500$ and $0.050$ when $n=250$; see the 5$th$ rows in Tables S4 and S7 in Section S2 of the Online Supplementary Material. We characterize rejection of $H_{0}$ under $H_{0}^{\lambda_1}$ as a type I error. In fact since the CIFs are determined by both $\lambda_{01}(t|X)$ and $\lambda_{02}(t|X)$, rejection of $H_{0}$ is a true positive finding in settings where $\beta^{\star} \not \approx 0$, even if it is not for the anticipated reason (i.e. a reduction in the $0-1$ intensity). The FG statistic tends to yield more powerful tests than the DB$_6$ statistic in settings most closely aligned with real-world applications.
    \vspace*{2mm}
\item If in secondary intensity-based analyses interest is in $H_{0}^{\lambda_1}$, analyses based on the log-rank test statistic $T^{\lambda_1}_{LR}$ or the Cox Wald statistic $T^{\lambda_1}_{Cox}$ are recommended and commonly used in practice. The performance of tests for $H_{0}^{\lambda_1}$ when $H_{0}$ is true is of less concern; the same applies to tests for $H_{0}^{\lambda_1 \lambda_2}$. A discussion of multiplicity adjustments is beyond our present scope. 
\end{enumerate}

\begin{landscape}
\begin{figure}[ht]
\centering
    \scalebox{0.67}{
    \includegraphics{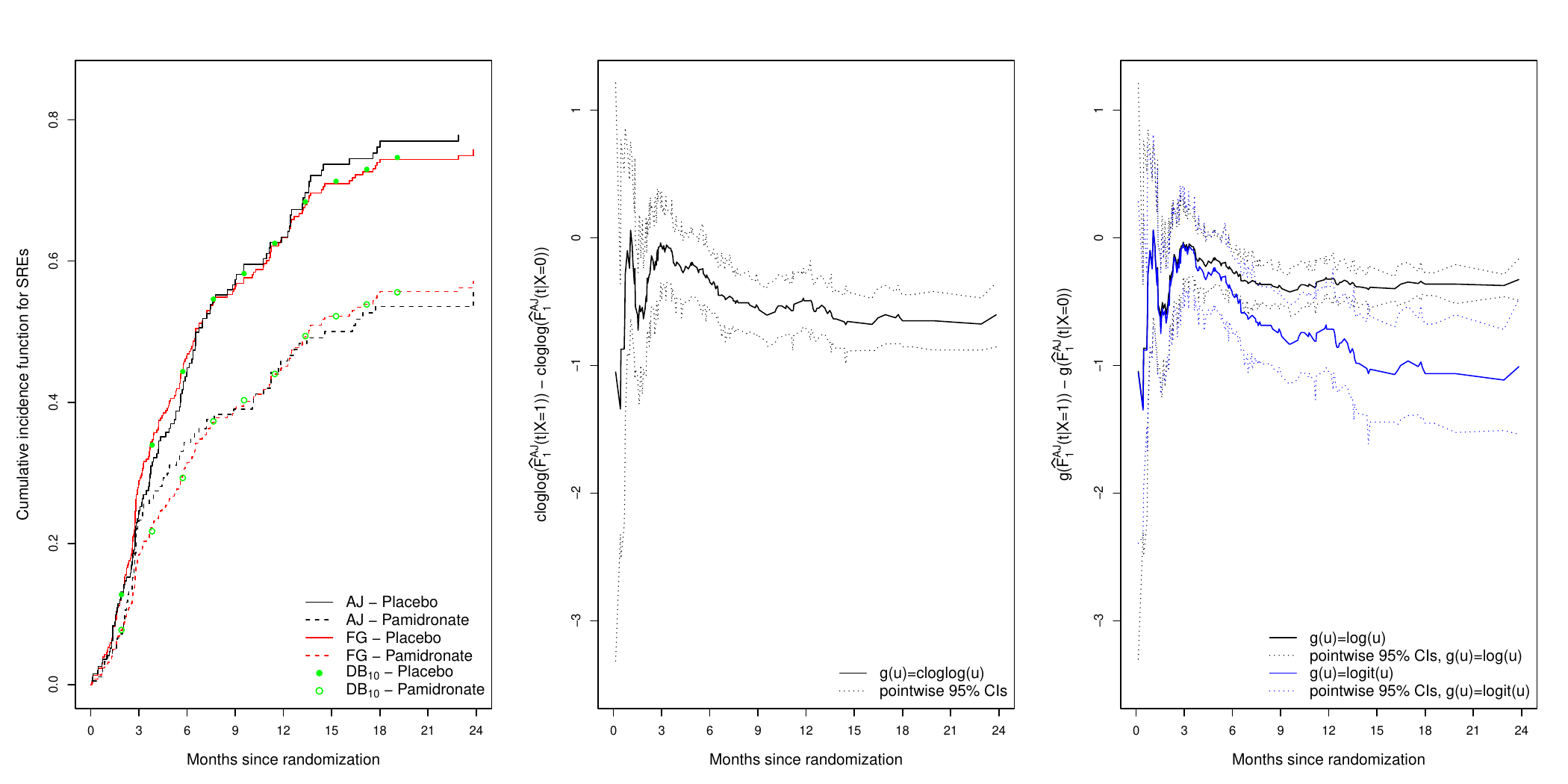}
    }
    \caption{Left panel: comparisons of nonparametric Aalen-Johansen (AJ) estimates of the cumulative incidence function for skeletal-related events (SREs) and semiparametric estimates implied by the Fine-Gray (FG) method and direct binomial (DB) estimation (with $R=10$ equi-spaced time points in $(0,21)$) under the cloglog transformation model (\ref{EventTime}), middle and right panels: plots of $g\bigl(\widehat{F}_1^{AJ}(t|X=1)\bigr)-g \bigl(\widehat{F}_1^{AJ}(t|X=0) \bigr)$ for $g(\cdot)=\text{cloglog}(\cdot), \log(\cdot)$ and $\text{logit}(\cdot)$ along with pointwise $95\%$ CIs; $X=1$: pamidronate group, $X=0$: placebo group.}
    \label{appl-F1versusF1hat}
\end{figure}
\end{landscape}

\section{Prevention of skeletal complications in metastatic breast cancer} \label{sec5} 
Here we consider analysis of data from an international multi-center randomized trial of women with stage IV breast cancer metastatic to bone, designed to assess the effect of a bisphosponate (pamidronate) on the prevention of skeletal complications including nonvertebral and vertebral fractures, the need for surgery to treat or prevent fractures, and the need for radiation for the treatment of bone pain \citep{Hortobagyi1996}.
Between January 1991 and March 1994 eighty-five sites in the United States, Canada, Australia and New Zealand recruited $380$ women, with $185$ randomized to receive a $90$ mg infusion of pamidronate every four week and $195$ to receive a placebo control. Each patient was followed until death, loss to follow-up or the administrative closure of the study. At the end of the 24 months follow-up period, $210$ women had experienced a skeletal complication, $33$ died complication-free, and $137$ were alive and complication-free. 
\newline
Our primary interest is the estimation of the effect of pamidronate on the cumulative incidence of skeletal complications, termed skeletal-related events (SREs). The Aalen-Johansen (AJ) estimates of $F_1(t|X)$ in the left panel of Figure \ref{appl-F1versusF1hat} (see black curves) show a steadily increasing cumulative probability of experiencing SREs in both treatment groups, with 2-year estimates of roughly $56 \%$ in the pamidronate group ($X$=1) and $78 \%$ in the placebo group ($X$=0); 
the two cumulative incidence curves begin to separate at 3 months. \cite{Gray1988}'s test of no treatment effect on the CIF for SREs gives a p-value of $0.000226$, and was carried out using the R function cuminc() in the competing risks package cmprsk. We note that this function also gives the nonparametric AJ estimate, but we obtained the CIF curves in Figure \ref{appl-F1versusF1hat} using the etm() function \citep{Allignol2011}. 

\noindent
Plots of $g\bigl(\widehat{F}^{AJ}_1(t|X=1)\bigr)-g\bigl(\widehat{F}^{AJ}_1(t|X=0)\bigr)$ along with pointwise $95\%$ CIs (constructed by the $\Delta$-method) can help to find an appropriate link function $g(\cdot)$. According to the middle and right panels of Figure \ref{appl-F1versusF1hat}, both the log and cloglog transformation offer reasonable approximations; the logit link approximates the differences in $\widehat{F}^{AJ}_1(t|X=1)$ and $\widehat{F}^{AJ}_1(t|X=0)$ less well. In the following analysis we restrict our attention to $g(u)=\text{cloglog}(u)$ and compare the FG and DB estimation procedures. We used the crr() function to fit model (\ref{EventTime}) under the FG procedure and comp.risk() for DB regression \citep{timereg-package}, as before in Section \ref{simulations-testing}. \newline \noindent
There was no evidence that censoring was dependent on covariates, so Kaplan-Meier (KM) estimates of a common censoring distribution were used in (\ref{equationFG1})-(\ref{equationFG2}) for the FG estimator and (\ref{ch3-DB-eqs}) for the DB estimator. The FG procedure yielded an estimated treatment effect of $\widehat{\beta}_{FG}=-0.516$ (robust SE=$0.141$, $95\%$ CI: $-0.793, -0.239$, p=$0.00026$), while DB estimation with $R=10$ equi-spaced time points over $(0,21)$ yielded $\widehat{\beta}_{DB}=-0.526$ (robust SE=$0.153$, $95\%$ CI: $-0.827, -0.226$, p=$0.000601$). There is not much difference in the estimated treatment effects, but a slight increase in the associated standard errors with the DB approach. The analysis suggests that pamidronate has a significant effect on the cumulative probability of SREs. The left panel of Figure \ref{appl-F1versusF1hat} also gives the estimated cumulative incidence curves obtained from fitting the semiparametric cloglog model (\ref{EventTime}) under each estimation procedure. Any difference between the semiparametric and the nonparametric estimates is an indirect reflection of model inadequacy. Both methods and the cloglog model offer a reasonable representation of the treatment effect, with some lack of fit early and later on in the trial. We complement the graphical model assessment with formal goodness-of-fit tests described in Section \ref{sec3.3}. Results can be found in Table \ref{gof-tests}, where the R package crrSC was used to perform the FG-based score test \citep{Zhou2013}. 
All other test statistics were calculated based on the results provided by crr() and comp.risk(). There is insufficient evidence to reject the assumption of a constant treatment effect over time, providing some justification for model (\ref{EventTime}).

\begin{table}[h]
\centering
\begin{tabular}{|clc|cc|}
\hline
\multicolumn{3}{|c|}{}                                                                                                                                       & \begin{tabular}[c]{@{}c@{}}Test\\ statistic\end{tabular} & p-value \\ \hline
\multicolumn{1}{|c|}{\multirow{4}{*}{FG}} & \multirow{2}{*}{Wald test of $H_{0}:\nu=0$}                            & $b(t)=t$                                & 2.444                                                    & 0.1180  \\
\multicolumn{1}{|c|}{}                    &                                                                        & $b(t)=\log(t)$                          & 1.967                                                    & 0.1608  \\ \cline{2-5} 
\multicolumn{1}{|c|}{}                    & \multirow{2}{*}{Score test of $H_{0}:\nu=0$}                           & $b(t)=t$                                & 2.690                                                    & 0.1010  \\
\multicolumn{1}{|c|}{}                    &                                                                        & $b(t)=\log(t)$                          & 2.148                                                    & 0.1427  \\ \hline
\multicolumn{1}{|c|}{\multirow{2}{*}{DB}} & Wald test of $H_{0}:\nu=0$                                             & $b(t)=t$                                & 2.262                                                    & 0.1326  \\
\multicolumn{1}{|c|}{}                    & \multicolumn{2}{l|}{Wald test of $H_{0}:\beta_r=\beta$ based on ($\ref{wald-DB}$) and $10000$ bootstrap samples} & 14.622                                                   & 0.1018  \\ \hline
\end{tabular}
\caption{Goodness-of-fit-tests for checking the assumption of a constant treatment effect over time.}
\label{gof-tests}
\end{table}

\begin{figure}[t]
\centering
\scalebox{0.52}{
\includegraphics{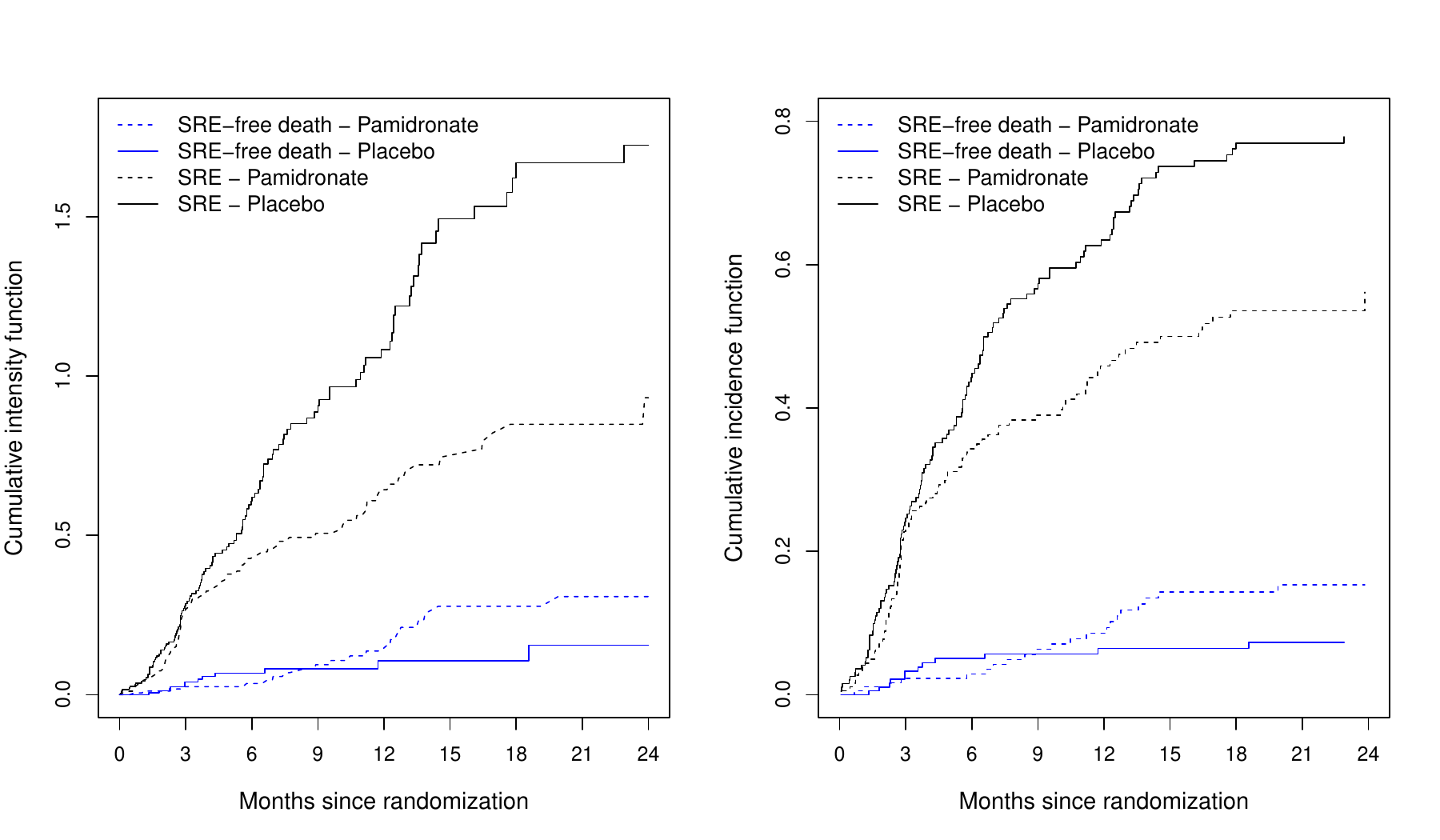}
}
\caption{Nelson-Aalen estimates of the cumulative intensity functions (left panel) and Aalen-Johansen estimates of the cumulative incidence functions (right panel) for skeletal complications and complication-free death for breast cancer patients in \citet{Hortobagyi1996}.}
\label{appl-fig1}
\end{figure}

Further secondary analyses were carried out to enhance understanding of the marginal treatment effects $\widehat{\beta}_{FG}$ and $\widehat{\beta}_{DB}$, including nonparametric and semiparametric analyses of the cause-specific hazard (CSH) functions for the time to SRE and death. Plots of the nonparametric Nelson-Aalen estimates of the CSH functions are displayed in the left panel of Figure \ref{appl-fig1} for each event type, and were obtained using the mvna package \citep{Allignol2008}. Within the first 3 months of follow-up pamidronate does not show any effect on the risk of SRE and death, which explains why $\widehat{F}_1^{AJ}(t|X=0) \approx \widehat{F}_1^{AJ}(t|X=1)$ and $\widehat{F}_2^{AJ}(t|X=0) \approx \widehat{F}_2^{AJ}(t|X=1)$ for $t \leq 3$ months; see right panel of Figure \ref{appl-fig1} and also (\ref{CIF}). Overall, patients treated with pamidronate have a significantly reduced hazard of skeletal complications (HR=$e^{\widehat{\gamma}_1}=0.590$, $95\%$ CI: $0.447, 0.778$, p=$0.000185$; log-rank test: p=$0.0002$) as compared to placebo patients; the results shown in the round brackets were obtained using the coxph() and survdiff() functions in the survival package \citep{survival-package}.
The risk of death, however, is slightly higher in the pamidronate arm; the crossing cause-specific hazards for complication-free death suggest that a Cox analysis would not be suitable in this case. The cumulative hazard in the placebo arm is relatively flat for complication-free death beyond 6 months, indicating that most complication-free deaths occurred in placebo patients within the first 6 months of follow-up. The secondary analyses demonstrate that the difference in the cumulative incidence curves for SREs is mainly due to the reduction in the intensity for SRE by the use of pamidronate.

\section{Discussion}
\label{sec6} 
Cumulative incidence function regression, which includes the FG model as a special case, has seen increasing use for the analysis of competing and semi-competing risks data in randomized clinical trials. Based on large sample results we have studied treatment effects in FG- and DB-based CIF regression when the true process has specified intensities, and established a robustness property for FG- and DB-based Wald tests of no treatment effect on $F_1(t|X)$ under key hypotheses concerning the intensity-based process. 
We focussed on the cloglog transformed CIF model, but other link functions can be likewise studied; see for instance \citet{Blanche2023} who recently investigated logistic CIF regression under binomial estimation. While DB estimation \citep{Scheike2008} can be used with any link function, the original \citet{Fine1999} procedure is defined in terms of the cloglog link. A key message in this paper is the need for secondary intensity-based analysis when assessing the effect of a randomized treatment on the CIF for the event of interest. We provide a brief summary of our main points and findings in the next paragraph. 
\newline \newline 
CIF regression targets a descriptive causal estimand based on an observable process feature \citep{Buehler2022}, but its interpretation is only meaningful if the underlying model assumptions are plausible. Both estimands $\beta_{FG}^\star$ and $\beta_{DB}^\star$ are functions of the degree of misspecification and the intensities of the underlying process including the treatment effects. 
We found that $\beta^{\star}$ has a simple interpretation if the incidence rate of the competing event is low and there is a relatively modest effect on the corresponding $0-2$ intensity, since in this case $\beta^{\star}$ is in close agreement with the treatment effect on the $0-1$ intensity.
In our numerical investigations the censoring distribution was assumed to be ``correctly" specified; in the case of a misspecified censoring model $\beta^{\star}$ is affected by the censoring process. It is straightforward to adapt the derivations of $\beta^{\star}$ given in Section \ref{sec3.1} to investigate how sensitive the limiting value is to such a misspecification. We refer to \citet{He2016} and \citet{Klein2005} who investigated in empirical studies the consequences of a misspecified censoring model on the estimation of $\beta$ when (\ref{EventTime}) is valid. \newline 
The robustness property in Theorem \ref{theorem2} supports the use of FG- and DB-based Wald tests using robust standard errors in primary analyses of clinical trials.
Similar robustness properties have been established for hazard-based Cox analyses in the simple survival setting \citep{DiRienzo2006} or logistic regression in binary response settings \citep{Rosenblum2009}.
\newline \newline \noindent
The planning stage of a trial, with primary estimation and testing based on model (\ref{EventTime}), should involve careful consideration of the process intensities, and how treatment may affect them. Control data collected in similar studies can offer insights into the nature of baseline intensities. Graphs such as those in Figure \ref{beta-star-FG-DB} can help to anticipate for $\widehat{\beta}$ under different assumed treatment effects in the true process. This could in turn be used for power and sample size calculations. According to the sample size formula for FG regression by \citet{Latouche2004}, the number of individuals $n$ required to ensure $100(1-\omega')\%$ power for a two-sided Wald test of $\beta=0$ vs ($\beta=\beta_1)$ at the $100\omega\%$ significance level is given by
\begin{align}
\label{samplesize-latouche}
n=\dfrac{1}{P(T_1 \leq C)} \cdot \dfrac{(z_{1-\omega/2}+z_{1-\omega'})^2}{\beta_1^2 P(X=1)P(X=0)} \; , 
\end{align}
where $P(T_1 \leq C)$ is the proportion of observed type $1$ events by the end of follow-up and $z_{q}$ is the $100q\%$ percentile of the standard normal distribution. Their Schoenfeld-type formula was derived under the simplistic assumption of ``complete data" (c.f. Section 3.1 of \citet{Fine1999}), in which case the FG at-risk indicator  
is known for all individuals and no inverse probability of censoring weighting in (\ref{equationFG1}) and (\ref{equationFG2}) is needed (i.e. $w_{i}(t)=1$ for all $t$ and $i=1,\ldots,n$). As a result, (\ref{samplesize-latouche}) was obtained by using a simple ``naive" model-based variance estimator for $\widehat{\beta}$ of the form $(P(X=1)P(X=0)N_{1\cdot}(\infty))^{-1}$ under both the null and alternative hypotheses, and so does theoretically not apply to trials where random right-censoring occurs. Simulation studies in \citet{Latouche2004} suggest that (\ref{samplesize-latouche}) provides a reasonable approximation in finite samples for a limited range of trial settings, provided the assumed model (\ref{EventTime}) is correct. Further study of the utility of the Latouche formula for more complex scenarios, e.g. when there is moderate to heavy right-censoring and the censoring intensity may depend on covariates, is however warranted. A closed-form sample size formula for DB regression is more challenging to derive and beyond the scope of this article. For now, we suggest a simulation-based approach to estimate power for DB-based Wald tests. \newline
The impact of non-proportional hazards in power calculations for the log-rank test has been extensively studied \citep{Zhang2009,Royston2016,Yung2020,Dormuth2023}. Similar investigations with FG and DB regression
have not yet received much attention. Large sample results such as those in (\ref{large-sample-FG-DB}) could facilitate sample size calculations for FG and DB regression under model misspecification.
\newline \newline \noindent
Similar issues to those discussed here arise with marginal \citet{Ghosh2002} and \citet{Mao2016} analyses in recurrent and terminal event settings (see e.g. Section 6.6 of \citealp{Cook2007}). 
The former model targets the marginal rate function for a potentially recurrent non-fatal event, recognizing that no further non-fatal event can occur after a fatal event. 
We again stress the need for care in interpreting treatment effects and secondary analyses for a fuller understanding of the response to treatment. Intensity-based Cox regression is recommended for assessing treatment effects on fatal and non-fatal events, and for assessing possible dependence of the fatal event intensity on the non-fatal event.
Rather than considering a one-dimensional estimand, \citet{Furberg2021c} proposed a two-dimensional estimand involving the expected number of non-fatal events via the Ghosh-Lin model and the survival function for the time to death through a Cox model of the form (\ref{sec2-Cox}), with estimation based on pseudo-observations \citep{Overgaard2023}.
In CV trials where patients may die from CV and non-CV related causes, their approach can be extended to a three-dimensional estimand by considering two separate FG models for the cumulative cause-specific death probabilities. The alternative Mao-Lin approach handles death from different causes by incorporating CV death into a composite recurrent event process \citep{Furberg2021}, which may be undesirable for interpretability. We study issues related to estimand specification in recurrent and terminal event settings in ongoing work \citep{buhler22}.
\newline \newline 
Other approaches have been suggested for processes with multiple events. One is utility-based methods in which utilities scores (costs or benefits) are specified for time spent in different states, or for specific events \citep{Torrance1987, Torrance1989, Gelber1989, Glasziou1990, Gelber1995, Cook2003}. 
A second is win ratio methods in which all possible sample paths (process histories) for an individual are given a (partial) ranking. Tests and estimates are then constructed by comparing all possible pairs $(i,j)$ and their outcomes $(\mathcal{H}_{i}(\infty), \mathcal{H}_{j}(\infty))$, where $X_{i}=1$ and $X_{j}=0$ \citep{Pocock2012, Oakes2016, Dong2020c, Mao2021, Yang2022}.
These and related approaches can be linked, and are discussed by \citet{Buehler2022}.

\bibliographystyle{apalike}
\bibliography{ref-paper2.bib}

\appendix

\section{Estimation of the censoring distribution and large sample results}
\label{appendix-censoring}
Let $V=(W, X)=((W_{1},W_{2},\ldots,W_{p-1}), X)'$ be a $p\times 1$ vector of observed baseline covariates that may be associated with the random censoring distribution so that $G(t|V)=P(C_r > t|V)$. We define $N_{r}(t)=\mathbbm{1}(C_r \leq t)$, and let $\{N_r(t),t>0\}$ denote the associated counting process for random loss to follow-up. If $\{Z(u), u>0\} \indep C_r \ | \ V$, then the censoring intensity is given by 
\begin{align*}
\lim_{\Delta t \downarrow 0} \dfrac{P(\Delta N_r(t) = 1 \ | \ C_r \geq t,  V)}{\Delta t} = \lim_{\Delta t \downarrow 0} \dfrac{P(\Delta N_r(t) = 1 \ | \ C_r \geq t, T \geq t, V)}{\Delta t} = \lambda_r(t | V) \; , 
\end{align*}
where $\Delta N_r(t)=N_r(t+\Delta t^-)-N_r(t^-)$ and $\lim_{\Delta t \downarrow 0} \Delta N_r(t)=dN_r(t)$. 
We further let the cumulative censoring intensity denote by $\Lambda_{r}(t|V)=\int_{0}^{t} d\Lambda_{r}(u|V)$ for $d\Lambda_{r}(u|V)=\lambda_{r}(u|V)du$. \newline 
We consider now a sample of $n$ independent individuals indexed by $i=1,2, \ldots,n$, and derive the limiting behaviour of Kaplan-Meier (KM) and Cox estimators of the censoring survivor function under different true censoring models.

\subsection{Limiting behaviour of the KM estimator under different true censoring mechanisms}
\label{cov-indep-cens}
\begin{enumerate}
\item If $C_{r}$ does truely not depend on covariates $V$, the censoring survivor function $G(t)=P(C_{r} > t)$ can be estimated by the Kaplan-Meier method where 
\begin{align} \label{ch3-KM-censoring}
\widehat{G}(t) = \prod_{u < t} \bigl( 1 - d\widehat{\Lambda}_r(u) \bigr) =  \prod_{u < t} \biggl( 1 - \dfrac{\sum_{i=1}^{n}Y_{i}(u)dN_{ri}(u)}{\sum_{i=1}^{n}Y_{i}(u)} \biggr)  \; , 
\end{align}
with $C_i=\min(C_{ri}, \tau), Y_{i}(t)=\mathbbm{1}(\min(T_i,C_i) \geq t)$, and the product taken over all unique random censoring times $u<t$. The Nelson-Aalen type estimate $d\widehat{\Lambda}_r(u)$ in (\ref{ch3-KM-censoring}) is obtained as the solution to 
$\sum_{i=1}^{n} Y_{i}(u) \bigl( dN_{ri}(u) - d\Lambda_{r}(u) \bigr)= 0,$
and is consistent for the true $d\Lambda_r(u)=\mathbb{E}(dN_{r}(u) | C_{r} \geq u)$ when the censoring process is Markov. For more general processes $d\widehat{\Lambda}_r(u)$ is consistent for the marginal censoring rate $d{\Lambda}^{\star}_r(u)=\mathbb{E}(dN_{r}(u) | C_{r} \geq u)$, the solution to $\mathbb{E}\bigl(\sum_{i=1}^{n} Y_{i}(u) (dN_{ri}(u) - d\Lambda_r(u) ) \bigr) = 0$ \citep{Cook2018}. Note that $d{\Lambda}^{\star}_r(u)=d{\Lambda}_r(u)$ for Markov processes. As a consequence,
\begin{align}
\label{Gstar-KM-1}
\widehat{G}(t) \longrightarrow G(t) &=  P(C_{r} > t; {d\Lambda}_r(\cdot)) = \prod_{u < t} \bigl( 1 - {d\Lambda}_r(u) \bigr) \; , \ \ \text{or} \\
\label{Gstar-KM-2}
\widehat{G}(t) \longrightarrow G^{\star}(t) &=  P(C_{r} > t; {d\Lambda}^{\star}_r(\cdot)) = \prod_{u < t} \bigl( 1 - {d\Lambda}^{\star}_r(u) \bigr) \; .
\end{align}

\item If the true censoring intensity $\lambda_r(t|V_{i})$ depends on $V_{i}=(W_{i}', X_{i})'$ but KM estimation stratified by $X_{i}$ only is used for estimation of the censoring survivor function, then 
\begin{align}
\widehat{G}(t|X_i) \longrightarrow G^{\star}(t |X_{i}) &=  P(C_{r} > t | X_{i}; {d\Lambda}^{\star}_r(\cdot | X)) \; , 
\end{align} 
where \begin{align*}
{d\Lambda}^{\star}_r(t | X) = \dfrac{\mathbbm{E}_{W}\bigl( \lambda_r(t|X,W)G(t|X,W)S(t|X,W)\bigr)}{\mathbbm{E}_{W}\bigl(G(t|X,W)S(t|X,W)\bigr)} \; ,   
\end{align*}
can be found as the solution to $\mathbb{E}\bigl(Y(u) \bigl( dN_{r}(u) - d\Lambda_r(u) \bigr) | \ X \bigr) = 0$ for $X=0,1$; with expectations taken with respect to the true censoring, competing risks and covariate processes. 
\end{enumerate}

\subsection{Limiting behaviour of the Cox estimator under misspecification} 
\label{cov-dep-cens}
\begin{enumerate}
\item The conditional censoring survivor function $G(t | V_i)=P(C_{ri} > t |V_i)$ can be estimated by 
\begin{align*}
\widehat{G}(t | V_i) = \prod_{u < t} \biggl( 1 - d\widehat{\Lambda}_{r}(u | V_i) \biggr)
\; , 
\end{align*}
where the censoring intensity may be modeled based on the Cox proportional hazards specification. If $S^{(l)}(u, \gamma_r)= \sum_{i=1}^{n} Y_{i}(u) V_{i}^{\otimes l} \exp(\gamma_r' V_{i})$ and $s^{(l)}(u, \gamma_r)=\mathbb{E}(S^{(l)}(u, \gamma_r))$ for $l=0,1$, the solution to 
\begin{align*}
U(\gamma_{r}) = \sum_{i=1}^{n}  \int_{0}^{\infty} Y_{i}(u) \biggl( V_{i} - \dfrac{S^{(1)}(u, \gamma_r)}{S^{(0)}(u, \gamma_r)} \biggr) dN_{ri}(u) = 0 
\end{align*}
gives the Cox partial likelihood estimator $\widehat{\gamma}_{r}$. The Breslow-type estimator of the cumulative baseline hazard is
$\widehat{\Lambda}_{r0}(t)=\int_{0}^{t} d\widehat{\Lambda}_{r0}(u)$, where $d\widehat{\Lambda}_{r0}(u)=\sum_{i=1}^{n} Y_{i}(u)dN_{ri}(u) \big / \sum_{i=1}^{n} Y_{i}(u)\exp(\widehat{\gamma}_r'V_i)$. Under conditionally independent censoring and a correctly specified censoring model, $\widehat{\gamma}_{r}$ is consistent for the true $\gamma_r$ and $d\widehat{\Lambda}_{r0}(t)$ for the true $d\Lambda_{r0}(t)$, in which case $$\widehat{G}(t|V_{i}) \longrightarrow G(t|V_{i})=P(C_{ri}>t \ | \ V_{i}; d\Lambda_{r0}(\cdot), \gamma_r) \; . $$
\item More generally, under some type of model misspecification, $\widehat{\gamma}_{r}$ is consistent for the limiting value $\gamma_{r}^{\star}$, the solution to
\begin{align*}
\mathbb{E} \bigl( U(\gamma_{r}) \bigr) = \int_{0}^{\infty} \biggl\{ \mathbb{E}\biggl(\sum_{i=1}^{n}  Y_{i}(u) V_{i} dN_{ri}(u) \biggr) - \dfrac{s^{(1)}(u, \gamma_r)}{s^{(0)}(u, \gamma_r)} \mathbb{E}\biggl( \sum_{i=1}^{n} Y_{i}(u)  dN_{ri}(u) \biggr) \biggr\} = 0 \; , 
\end{align*}
where expectations are taken with respect to the true censoring, covariate and competing risks processes \citep{Struthers1986, Lin1989}.
If the probability limit of $\widehat{\Lambda}_{r0}(u)$ is denoted by $d\Lambda_{r0}^{\star}(u)$, we find that 
\begin{align}
\label{Gstar-Cox}
\widehat{G}(t|V_{i}) \longrightarrow G^{\star}(t|V_{i}) = P(C_{ri} > t \ | \ V_{i}; d\Lambda^{\star}_{r0}(\cdot), \gamma_r^{\star}) \; .
\end{align}
\end{enumerate}

\section{Proof of Theorem 2}
\label{app-proof-theorem2}
Here we prove the robustness property of Theorem \ref{theorem2} for both estimation methods; see Section \ref{proof-FG} for the FG procedure and Section \ref{proof-DB} for the DB procedure. Specifically, we show that $\beta_{FG}^{\star}=0$ and $\beta_{DB}^{\star}=0$ under $H^{\lambda_1\lambda_2}_0$ and $H_{0}$, but $\beta_{FG}^{\star} \neq 0$  and $\beta_{DB}^{\star} \neq 0$ under $H_{0}^{\lambda_1}$. 

\subsection{Proof of the robustness property for the FG-based Wald test}
\label{proof-FG}
We assume $G^{\star}=G$, and let $s^{(l, \star)}(t, \beta)=s^{(l)}(t, \beta)=\mathbb{E}_{X}\bigl((1-F_1(t|X))X^l \exp(\beta X)\bigr)$ and $s^{(l, \star)}(t)=s^{(l)}(t)=\mathbb{E}_{X}\bigl(f_1(t|X)X^l\bigr)$ for $l=0,1$. Consider then (\ref{ch3-limitingFG}) labelled as $\mathbb{E}(U^{FG})$ for the remainder of this section: 
\begin{align}
\nonumber \mathbb{E}(U^{FG}) &= \mathbb{E}(U^{FG}(\beta, G^{\star})) = \int_{0}^{\infty} \biggl \{ s^{(1)}(t) - \dfrac{s^{(1)}(t, \beta)}{s^{(0)}(t, \beta)} s^{(0)}(t)\biggr \} dt \\
\label{FG-eq1-proof-th2} &=\int_{0}^{\infty} \biggl \{ \mathbb{E}_{X}(Xf_1(t|X)) - \dfrac{\mathbb{E}_{X}((1-F_1(t|X))Xe^{\beta X})}{\mathbb{E}_{X}((1-F_1(t|X))e^{\beta X})} \mathbb{E}_{X}(f_1(t|X)) \biggr \} dt \; .
\end{align}
\begin{enumerate}
\item \underline{Under $H_{0}^{\lambda_1\lambda_2}$:} \ \ $f_1(t|X)=f_1(t), F_1(t|X)=F_1(t)$
\begin{align}
\nonumber \therefore \ \mathbb{E}(U^{FG}) &= \int_{0}^{\infty} f_1(t) \biggl \{ \mathbb{E}_{X}(X) - \dfrac{\mathbb{E}_{X}(X e^{\beta X})}{\mathbb{E}_{X}(e^{\beta X})} \biggr \} dt 
= P(T_1 < \infty) \biggl \{ \mathbb{E}_{X}(X) - \dfrac{\mathbb{E}_{X}(Xe^{\beta X})}{\mathbb{E}_{X}(e^{\beta X})} \biggr \} \\
\label{FG-eq-proof-th2} & \overset{!}{=} 0 \ \ \Longleftrightarrow \ \ \dfrac{\mathbb{E}_{X}(Xe^{\beta X})}{\mathbb{E}_{X}(e^{\beta X})} = \mathbb{E}_{X}(X), 
\end{align}
which holds iff $\beta=0$. \newline \newline
\hspace*{0.3cm} \textit{To show}: \ \ $\mathbb{E}_{X}(Xe^{\beta X}) = 0 \ \ \Longleftrightarrow \ \ \beta=0$. \newline
\hspace*{0.3cm} \textit{Proof}: \ \ We assume $\mathbb{E}_{X}(X)=0$ without loss of generality.  \newline 
\hspace*{0.3cm} $''\Longleftarrow''$: \ \ If $\beta=0$, then $\mathbb{E}_{X}(Xe^{\beta X}) = \mathbb{E}_{X}(X)=0$. \newline 
\hspace*{0.3cm} $''\Longrightarrow''$: \ \ Let $l(x) \geq 0$ be the probability density function of $X$ and assume $\beta \neq 0$. Since 
$$ \int_{-\infty}^{0} x e^{\beta x}l(x)dx = \int_{0}^{\infty}x e^{-\beta x}l(-x)dx \; , $$
\hspace*{0.3cm} we find that 
\begin{align}
\nonumber \mathbb{E}_{X}(Xe^{\beta X}) &= \int_{-\infty}^{\infty} xe^{\beta x}l(x)dx = \int_{-\infty}^{0} xe^{\beta x}l(x)dx + \int_{0}^{\infty} xe^{\beta x}l(x)dx \\
\label{eq-proof} &= \int_{0}^{\infty} x \bigl(\underbrace{e^{-\beta x}}_{>0 \ \text{if} \ \beta \ne \ 0}l(-x) + \underbrace{e^{\beta x}}_{>0 \ \text{if} \ \beta \ne \ 0}l(x) \bigr) dx \; . 
\end{align}
\hspace*{0.3cm} If $\beta \neq 0$ the integrand in (\ref{eq-proof}) is $>0$ and, consequently, $\mathbb{E}_{X}(Xe^{\beta x})>0$. By proof of contradiction, $\beta$ must be equal to $0$.  
\newline
\item \underline{Under $H_{0}$:} \ \ This gives (\ref{FG-eq-proof-th2}) also. \newline 
\item \underline{Under $H_{0}^{\lambda_1}$:} \ \ In this case $f_1(t|X)=\lambda_{01}(t)S(t|X)=\lambda_{01}(t)e^{-\Lambda_{01}(t)-\Lambda_{02}(t|X)}, F_1(t|X)=\int_{0}^{t}\lambda_{01}(u)S(u|X)du$, and (\ref{FG-eq1-proof-th2}) $\neq 0$ unless $\lambda_{02}(t|X)=\lambda_{02}(t)$.
\end{enumerate}
\subsection{Proof of the robustness property for the DB-based Wald test}
\label{proof-DB}
We assume $G^{\star}=G$, and consider (\ref{ch3-limitingDB-1}) and (\ref{ch3-limitingDB-2}) labelled as $\mathbb{E}(U^{DB}_{\alpha_r})$ and $\mathbb{E}(U^{DB}_{\beta})$ for the remainder of this section:  
\begin{align*}
&\mathbb{E}(U^{DB}_{\alpha_r}) = \mathbb{E}(U_1(\alpha_r, G^{\star})) = \mathbb{E}_{X} \biggl( \dfrac{h'(\alpha_r+\beta X)\bigl(F_1(s_r | X) - h(\alpha_r + \beta X)\bigr)}{h(\alpha_r+\beta X)(1-h(\alpha_r+\beta X))} \biggr) \\
&\mathbb{E}(U^{DB}_{\beta}) = \mathbb{E}(U_1(\beta,  G^{\star})) =  \sum_{r=1}^{R} \mathbb{E}_{X} \biggl(  \dfrac{X \cdot h'(\alpha_r+\beta X) \bigl( F_1(s_r | X) - h(\alpha_r + \beta X) \bigr)}{h(\alpha_r+\beta X)(1-h(\alpha_r+\beta X))} \biggr) \; . 
\end{align*}
\begin{enumerate}
\item \underline{Under $H_{0}^{\lambda_1\lambda_2}$:} \ \ $F_1(s_r|X)=F_1(s_r)$ for all $r=1,2,\ldots,R$
\begin{align}
\nonumber \therefore \ \mathbb{E}(U^{DB}_{\alpha_r}) &= F_1(s_r) \mathbb{E}_{X} \biggl( \dfrac{h'(\alpha_r+\beta X)}{h(\alpha_r+\beta X)(1-h(\alpha_r+\beta X))} \biggr) - \mathbb{E}_{X} \biggl( \dfrac{ h'(\alpha_r+\beta X)}{(1-h(\alpha_r+\beta X))} \biggr) \\
\nonumber &= \underbrace{{\dfrac{P(X=0)h'(\alpha_r)}{(1-h(\alpha_r))}}}_{>0} \biggl(\dfrac{F_1(s_r)}{h(\alpha_r)} - 1 \biggr) + \underbrace{{\dfrac{P(X=1)h'(\alpha_r+\beta)}{(1-h(\alpha_r+\beta))}}}_{>0} \biggl( \dfrac{F_1(s_r)}{h(\alpha_r+\beta)} - 1 \biggr) \\
\label{db-eq1-proof-th2} &\overset{!}{=} 0 \ \ \Longleftrightarrow \ \ \ F_1(s_r)=h(\alpha_r) \ \text{and} \ F_1(s_r)=h(\alpha_r+\beta) \ \ \text{for all} \ r=1,2,\ldots,R \; , 
\end{align}
since $h(u)=g^{-1}(u)=1-\exp(-\exp(u))  >0$ and $h'(u)=\exp(u-\exp(u)) >0$ for all $u$. Similarily,
\begin{align}
\nonumber \mathbb{E}(U^{DB}_{\beta}) &= \sum_{r=1}^{R} \biggl \{ F_1(s_r)  \mathbb{E}_{X} \biggl( \dfrac{X \cdot h'(\alpha_r+\beta X)}{h(\alpha_r+\beta X)(1-h(\alpha_r+\beta X))} \biggr) - \mathbb{E}_{X} \biggl( \dfrac{X \cdot h'(\alpha_r+\beta X)}{(1-h(\alpha_r+\beta X))} \biggr) \biggr \} \\
\nonumber &= \sum_{r=1}^{R} \underbrace{{\dfrac{P(X=1)h'(\alpha_r+\beta)}{(1-h(\alpha_r+\beta))}}}_{>0} \biggl( \dfrac{F_1(s_r)}{h(\alpha_r+\beta)} - 1 \biggr)  \\
\label{db-eq2-proof-th2} &\overset{!}{=} 0 \ \ \Longleftrightarrow \ \ F_1(s_r)=h(\alpha_r+\beta) \ \ \text{for all} \ r=1,2,\ldots,R \; .
\end{align}
Note that $\beta=0$ solves (\ref{db-eq1-proof-th2}) and (\ref{db-eq2-proof-th2}). 
\newline
\item \underline{Under $H_{0}$:} \ \ This gives (\ref{db-eq1-proof-th2}) and (\ref{db-eq2-proof-th2}) also. 
\newline
\item \underline{Under $H_{0}^{\lambda_1}$:} \ \ In this case (\ref{db-eq1-proof-th2}) and (\ref{db-eq2-proof-th2}) do not hold. 
\end{enumerate}

\end{document}